\documentclass[12pt]{article}
\pdfoutput=1

\usepackage{putex}

\usepackage{cancel}
\usepackage{simplewick}
\usepackage{multirow}
\usepackage{comment}

\usepackage{graphicx}
\usepackage{caption}
\usepackage{amsmath}
\usepackage{array}
\usepackage{subcaption}
\usepackage{epstopdf}
\usepackage{enumerate}
\usepackage{cite}
\usepackage{youngtab}
\usepackage{tensor}
\usepackage{slashed}
\usepackage[aligntableaux=center]{ytableau}
\usepackage[utf8]{inputenc}
\usepackage{rotating}
\usepackage{bigfoot}

\usepackage[
      colorlinks=true,
      linkcolor=blue,
      urlcolor=blue,
      filecolor=black,
      citecolor=red,
      ]{hyperref}

\usepackage[compat=1.1.0]{tikz-feynman}
\usetikzlibrary{decorations.markings}

\numberwithin{equation}{section}

\def \( {\left(}
\def \) {\right)}
\def \< {\left<}
\def \> {\right>}
\def \eps {\varepsilon}

\newcommand{\be}{\begin{equation}} \newcommand{\ee}{\end{equation}}

\newcommand\tpind[5]{\langle {\cal O}_{#1}^{#4}\left(#2\right) {\cal O}_{#1}^{#5}\left(#3\right) \rangle}
\newcommand\tp[3]{\tpind{#1}{#2}{#3}{}{}}

\newcommand\thpnormind[5]{\langle {\cal O}_{\Delta_0}\left(#1\right) \hat{\cal O}_{\Delta_0}\left(#2\right) {\cal O}_{#3}^{#5}\left(#4\right) \rangle}
\newcommand\thpnorm[4]{\thpnormind{#1}{#2}{#3}{#4}{}}
\newcommand\thptildind[5]{\langle {\cal O}_{\tilde\Delta_0}\left(#1\right) \hat{\cal O}_{\tilde\Delta_0}\left(#2\right) {\cal O}_{#3}^{#5}\left(#4\right) \rangle}
\newcommand\thptild[4]{\thptildind{#1}{#2}{#3}{#4}{}}

\tikzfeynmanset{
	O/.style = {very thick}
}
\tikzfeynmanset{
	Ot/.style = {scalar}
}
\tikzfeynmanset{
	Th/.style = {double distance=2pt}
}
\tikzfeynmanset{
	dO/.style = {fermion}
}

\newcommand{\es}[2] {\begin{equation} \label{#1} \begin{split} #2 \end{split} \end{equation}}

\begin{document}

\institution{WZ}{Department of Particle Physics and Astrophysics, Weizmann Institute of Science, Rehovot, Israel}

\title{AdS from CFT for scalar QED}

\authors{Ofer Aharony\worksat{\WZ}\footnote{e-mail: {\tt ofer.aharony@weizmann.ac.il}}, Shai M.~Chester\worksat{\WZ}\footnote{e-mail: {\tt iahs81@gmail.com}} and Erez Y.~Urbach\worksat{\WZ}\footnote{e-mail: {\tt erez.urbach@weizmann.ac.il}}}

\abstract{
We construct an explicit bulk dual in anti-de Sitter space, with couplings of order $1/N$, for the $SU(N)$-singlet sector of QED in $d$ space-time dimensions ($2 < d < 4$) coupled to $N$ scalar fields. We begin from the bulk dual for the theory of $N$ complex free scalar fields that we constructed in our previous work, and couple this to $U(1)$ gauge fields living on the boundary in order to get the bulk dual of scalar QED (in which the $U(1)$ gauge fields become the boundary value of the bulk vector fields). As in our previous work, the bulk dual is non-local but we can write down an explicit action for it. We focus on the CFTs arising at low energies (or, equivalently, when the $U(1)$ gauge coupling goes to infinity). For $d=3$ we discuss also the addition of a Chern-Simons term for $U(1)$, modifying the boundary conditions for the bulk gauge field. We also discuss the generalization to QCD, with $U(N_c)$ gauge fields coupled to $N$ scalar fields in the fundamental representation (in the large $N$ limit with fixed $N_c$).
}
\date{September 2021}

\maketitle

\tableofcontents

\section{Introduction and Summary}
\label{intro}

The AdS/CFT correspondence \cite{Maldacena:1997re,Witten:1998qj,Gubser:1998bc} is an equivalence between theories of quantum gravity on asymptotically anti-de Sitter (AdS) space in $d+1$ dimensions, and conformal quantum field theories (CFTs) in $d$ dimensions. Since for $d \geq 2$ we do not have any non-perturbative definition of these quantum gravity theories, the correspondence should be viewed as providing a non-perturbative definition for these theories in terms of the corresponding quantum field theories, whenever those are known and well-understood. In many cases of the correspondence the gravitational theories have a classical limit, in which the ratio of the Planck scale to the radius of anti-de Sitter space goes to infinity, and which corresponds to a large $N$ limit of some sort in the CFT side. In those cases one can test the correspondence by checking that the semi-classical expansion on the gravity side agrees with the large $N$ limit of the corresponding field theories.

A derivation of the AdS/CFT correspondence requires showing that the CFT can be rewritten as a quantum gravity theory, in the sense that its $1/N$ expansion reproduces the corresponding perturbative expansion around some gravitational solution in AdS space. While one can often test this for specific observables, providing such a derivation for the full theory is challenging (see \cite{Eberhardt:2019ywk,Dei:2020zui,Eberhardt:2020akk,Bertle:2020sgd} for some recent progress in relating free $d=2$ symmetric orbifold CFTs to string theory on $AdS_3$). In our recent paper \cite{Aharony:2020omh} we provided such a derivation for the simplest case of the AdS/CFT correspondence -- the duality \cite{Klebanov:2002ja} between the $U(N)$-singlet sector of $N$ free complex scalar fields in $d$ dimensions, and a theory of higher-spin gravity on $AdS_{d+1}$. Our derivation in \cite{Aharony:2020omh} followed the methods of bi-local holography \cite{Das:2003vw,Koch:2010cy,Koch:2014aqa,Koch:2014mxa,deMelloKoch:2012vc,deMelloKoch:2018ivk}, in its Lorentz-invariant version. We first changed variables in the CFT from the $N$ free scalar fields $\phi_I(x)$ ($I=1,\cdots,N$) to their bi-local combinations
\be \label{defg}
G(x_1,x_2) \equiv {1\over N} \sum_{I=1}^N \phi_I^*(x_1) \phi_I(x_2),
\ee
which capture all the information about $U(N)$-invariant observables in the theory (at least in flat space). In the large $N$ expansion one can write down an action for $G(x_1,x_2)$ that reproduces the correlators of the original theory, to all orders in $1/N$ (as discussed in \cite{Aharony:2020omh}, the mapping to the bi-local variables makes sense also for finite $N$ but it is more subtle then, and in this paper we only discuss the theory in the $1/N$ expansion). In order to map this to a theory in AdS space, we first expanded the bi-local field $G(x_1,x_2)$ in a basis of eigenstates of the Euclidean conformal group $SO(d+1,1)$. We then showed that the same eigenstates appear in the expansion of transverse traceless fields $\Phi_J(x,z)$ of spin $J$ ($J=0,1,2,\cdots$) living on a fixed $AdS_{d+1}$ space, enabling a one-to-one mapping between these fields and the bi-local \eqref{defg}. Using this mapping we rewrote the action of the original theory as an action (with coupling $1/N$) for fields in AdS space. This action is explicitly known but is non-local. For a specific choice of the undetermined coefficients appearing in the mapping, the quadratic term in the bulk can be chosen to be local. Since the $U(N)$-singlet sector of $N$ free scalars is conjectured \cite{Klebanov:2002ja} to be dual to Vasiliev's high-spin gravity theory \cite{Vasiliev:1990en,Vasiliev:1992av,Vasiliev:1995dn}, we believe that our action is equivalent to a gauge-fixed version of this theory, in which its fields live on a fixed AdS space-time (as in \cite{Neiman:2015wma}); the spectrum of physical fields is consistent with this, but the equivalence has not yet been shown.

In this paper, we generalize this construction to the $SU(N)$-singlet sector of scalar QED -- the theory of a $U(1)$ gauge field minimally coupled to $N$ charged scalar fields, for $2 < d < 4$. In principle this generalization is straightforward, since on the field theory side it just involves introducing a dynamical $U(1)$ gauge field and coupling it to $SU(N)$-singlet combinations of the scalar fields, and we can in principle do this also on the gravity side. On-shell, local $SU(N)$-invariant operators (such as the ones appearing in the coupling to the gauge field) map to the boundary values of the bulk fields on AdS space; however, the construction above requires the off-shell mapping of the local CFT operators to the bulk, and at first sight this seems to be more complicated. In \cite{Aharony:2020omh} we showed that for $J=0$ and $d<4$ the naive on-shell mapping works also off-shell, and we used it there to derive the dual to the critical $U(N)$ (or $O(N)$) vector model, by coupling a singlet $\sigma(x)$ to the spin $J=0$ operator $\phi_I(x) \phi_I(x)$. In this paper we show that the same is true also for $J=0$ and $d>4$, and also for higher values of $J$. This allows us to write down a simple action for the $d$-dimensional $U(1)$ gauge field coupled to the bulk fields; in fact, as may have been expected \cite{Klebanov:1999tb,Witten:2001ua,Witten:2003ya,Chang:2012kt,Berkooz:2002ug}, the $U(1)$ gauge field becomes simply the boundary value of the $(d+1)$-dimensional vector field $\Phi_1(x,z)$. Thus, compared to the original theory of $N$ free scalars, the description of QED just involves changing the boundary condition for the bulk vector field $\Phi_1$, and perhaps adding some boundary terms for it. The boundary terms are required if one wants to describe QED at finite coupling, but they disappear in the low-energy limit, where QED (for $2 < d < 4$) flows to a conformal field theory (at least at large $N$ and to all orders in $1/N$  \cite{Appelquist:1988sr}, and perhaps also for all finite values of $N$ \cite{Nguyen_1999,Kajantie:2004vy}). In the special case of $d=3$ one can also add a Chern-Simons coupling at level $k$; when $k/N$ is kept fixed in the large $N$ limit, this coupling is translated into a modified boundary condition for the bulk vector field.

One application of our bulk formalism for QED would be to find classical solutions in the bulk describing monopole operators, and perhaps to use them to compute their correlation functions. As we discuss below, our formalism only includes monopole operators if there is no Chern-Simons coupling for the $U(1)$, and it may be interesting to look for ways to get around this problem so that monopoles in Chern-Simons-matter theories may also be incorporated.

We also generalize our construction to the $SU(N)$ singlet sector of $N$ scalar fields in the fundamental representation of a $U(N_c)$ gauge field (``scalar QCD" \cite{PhysRevLett.64.721}), when $N_c$ is kept fixed in the large $N$ limit, and for $d=3$ we can again include a Chern-Simons level $k$. The bulk dual is very similar to the $N_c=1$ QED case, except the bulk fields now transform in the adjoint representation of $U(N_c)$. Note that this limit is different from the large $N_c,k$ and fixed $\lambda\equiv N_c/k$ limit considered in \cite{Aharony:2011jz}, which is believed to also be dual to a parity breaking higher spin bulk theory. It would be interesting to see if the generalization of our construction to this limit just involves changing boundary conditions, or explicitly changes the bulk interaction terms as a function of $\lambda$ (see discussions in \cite{Giombi:2012ms,Chang:2012kt}). Such a generalization, as well as the inclusion of fermions, would allow us to similarly construct the bulk dual of ABJ theory \cite{Aharony:2008gk} in its higher spin limit, which could ultimately lead to a connection to string theory and M-theory via the ABJ triality \cite{Chang:2012kt,Binder:2021cif}.

Finally, our off-shell mapping allows us to construct the bulk dual for the critical $U(N)$ or $O(N)$ vector model for $4<d<6$, as discussed in \cite{Fei:2014yja,Chester:2014gqa}. This CFT is not unitarity at finite $N$ due to complex large $N$ instantons \cite{Giombi:2019upv}, but is unitary in the $1/N$ expansion. It would be interesting to study the classical solutions corresponding to these instantons from the bulk perspective using our construction.

We begin in section \ref{sec:QED_bilocal} by describing the $SU(N)$-singlet sector of scalar QED in bi-local variables, and translating its action to these variables. In section \ref{sec:map} we review the mapping found in  \cite{Aharony:2020omh} between the free scalar theory and the bulk, and extend it to an off-shell mapping for local operators in the CFT. In section \ref{sec:QED_bulk} we then use this mapping to construct the bulk dual for the $SU(N)$-singlet sector of scalar QED, and show that it reproduces the $1/N$ expansion of this theory. Finally, in section \ref{sec:QCD_bulk} we discuss the generalization to scalar QCD.

\section{Scalar QED in the bi-local formalism}
 \label{sec:QED_bilocal}
 
 We begin in this section by discussing QED with $N$ charged scalars in the bi-local formalism. First we will discuss scalar QED in the usual local formalism in a general dimension $2<d<4$, as well as in $d=3$ where we can add a $U(1)$ Chern-Simons term with level $k$. We will then review how in $d=3$ one can realize the restriction to $SU(N)$-singlet fields in a local fashion, by coupling to a non-Abelian $U(N)$ Chern-Simons theory at level $k'$ and taking $k'\to\infty$, before coupling to the $U(1)$ gauge field of QED. Finally, we will write the path integral for the $SU(N)$-singlet sector of scalar QED in the bi-local language, following the discussion of free scalars in \cite{Das:2003vw,Jevicki:1979mb}, and perform a saddle point expansion at large $N$. In the 3d case we fix $\kappa=k/N$ to be finite in the large $N$ limit. 
 
 \subsection{The local formalism}
 \label{sec:local_QED}
 
 We begin by discussing the conventional local formalism for QED with $N$ complex scalars $\phi^I$ on Euclidean $\mathbb{R}^d$ for $2<d<4$, following \cite{Chester:2016ref,Chester:2017vdh,Chester:2021drl,Giombi:2015haa}. The actions for this theory can be assembled by starting with the (Euclidean) free theory action
 \es{freeS}{
  \mathcal{S}_\text{free}[\phi]=\int d^dx\partial_i\phi^*_I(x) \partial_i\phi^I(x)\,,
 }
 where all repeated indices are summed, including the upper/lower $I=1,\cdots,N$ indices for the fundamental/anti-fundamental of the $U(N)$ global symmetry, and the $i=1,\cdots,d$ index for the $SO(d)$ Euclidean rotations. We can then gauge the $U(1)$ subgroup by adding a Maxwell term with scalar couplings:
 \es{gaugeS}{
  \mathcal{S}_e[\phi,A]=\int d^dx\Bigg[\frac{1}{4e^2}F_{ij}F_{ij}+J_i[\phi]\frac{A_i}{\sqrt{N}}+\frac{A_i A_i}{N}\phi_I^*\phi^I\Bigg]\,,\quad J_i[\phi]\equiv i(\phi_I^*\partial_i\phi^I-\phi^I\partial_i\phi_I^*)\,,
 }
 where $A_i$ is a $U(1)$ gauge field with field strength $F_{ij}$, the $U(1)$ current is  $J_i$, the seagull term is required by gauge-invariance, and we rescaled $e^2\to e^2/N$ and $A_i \to A_i /\sqrt{N}$ for later convenience in the large $N$ limit. This action retains the $SU(N)$ subgroup of the original $U(N)$ global symmetry, and also has a new $(d-3)$-form $U(1)_T$ global symmetry generated by the Hodge dual field strength $* F$ \cite{Kapustin:2005py,Chester:2015wao}. The $\phi^I$ are uncharged under $U(1)_T$, but we can couple it to codimension-3 operators, which in $d=3$ become local operators called monopole operators \cite{Borokhov:2002ib}. Finally, we should add a quartic coupling $\frac{\lambda}{4N}(\phi^*_I\phi^I)^2$, which can be equivalently written by using a Hubbard-Stratonovich field $\sigma$ as 
 \es{HSS}{
 \mathcal{S}_{\lambda}[\phi,\sigma]=\int d^dx\Bigg[-\frac{1}{4\lambda}\sigma^2+\frac{1}{2\sqrt{N}}\sigma\phi_I^*\phi^I\Bigg]\,,
 }
where integrating out  $\sigma$ recovers the  $\frac{\lambda}{4N}(\phi^*_I\phi^I)^2$ term. The most general UV action for a $U(1)$ gauge theory coupled to $N$ complex scalars with relevant couplings in $2<d<4$ is then given by $ \mathcal{S}_\text{free}+ \mathcal{S}_e+\mathcal{S}_{\lambda}$, as well as a mass term (which we will always fine-tune to have vanishing mass at low energies). This theory is believed to flow in the IR to an interacting CFT called the $\mathbb{CP}^{N-1}$ or Abelian Higgs model\footnote{This expectation is supported by lattice data for $d=3$ and all values of $N$ \cite{Nguyen_1999,Kajantie:2004vy}, and by analytic results in the large $N$ limit for all values of $d$ \cite{Appelquist:1988sr}. We will take \eqref{CPNS} to be a formal definition of the IR CFT, which should be valid for all values of $d$ and $N$ for which this CFT exists.}:
\es{CPNS}{
\mathcal{S}_{\mathbb{CP}^{N-1}}[\phi,A,\sigma]\equiv\mathcal{S}_\text{free}[\phi]+ \mathcal{S}_{e\to\infty}[\phi,A]+\mathcal{S}_{\lambda\to\infty}[\phi,\sigma]\,,
}
 where $e,\lambda\to\infty$ since the Maxwell term is irrelevant while $(\phi^*_I\phi^I)^2$ is relevant (near the free UV fixed point).\footnote{At large $N$ one can also tune $\lambda=0$ to get a different IR CFT called the tricritical QED theory, but for $d=3$ it is not clear if the fixed point exists at finite $N$.} 
In $d=3$, we can generalize this CFT by adding a level $k$ Chern-Simons term
 \es{CSaction}{
 \mathcal{S}_{k}[A]=-\int d^3x\frac{ik}{4\pi N}\varepsilon_{ijl}A_i\partial_j A_l\,,
 }
where $k\in\mathbb{Z}$ (recall that we rescaled the gauge field by a factor of $\sqrt{N}$). Note that this term is marginal in 3d, unlike the Maxwell term which dropped out for $e\to\infty$.
 
\begin{figure}[t]
\centering
   \includegraphics{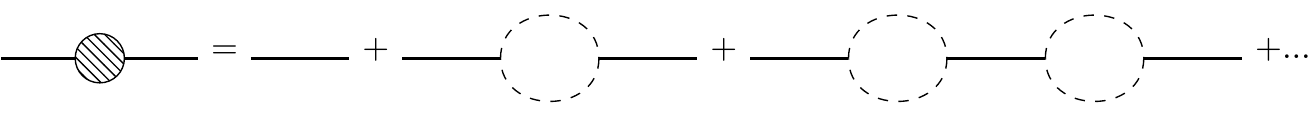}
   \caption{Bubble diagrams contributing to the effective $\sigma$ propagator $\langle\sigma(x)\sigma(0)\rangle_\lambda$ at leading $O(N^0)$ order. Thick lines correspond to $\sigma$ contractions, and dashed lines to the scalars $\phi_I$.}
   \label{fig:bubbles1}
\end{figure}

In general the fixed point written above is strongly coupled, but it becomes weakly coupled at large $N$, and it can be studied perturbatively in a $1/N$ expansion. We begin by writing down the Feynman rules for this expansion.
The propagator for $\phi^I$ is
 \es{G0}{
 G_0(x_1,x_2)=\frac{\Gamma(d/2-1)}{4\pi^{d/2}}\frac{1}{|x_1-x_2|^{d-2}}\,.
 }
For $\sigma$, we have the momentum space $\sigma$ propagator and the one-loop correction from the $\phi^2$ bubble diagram
\es{sigBubble}{
D_{\sigma}(p)=-2\lambda\,,\qquad \mathbb{B}_{\phi^2}(p)= -|p|^{d-4}\frac{
	2^{1-2d}\pi^{\frac{3-d}{2}}}{\Gamma(\frac{d-1}{2})\sin(\frac{\pi d}{2})}\,,
}
where the former comes from the trivial quadratic term $\sigma^2$, while the latter comes from the $\frac{1}{2\sqrt{N}}\sigma \phi_I^*\phi^I$ vertex and is simply $G_0(x,0)^2/4$ in momentum space. Since both are $O(N^0)$, at leading order in $1/N$ we need to resum an infinite geometrical series of bubble diagrams to obtain the effective $\sigma$ propagator (see Figure \ref{fig:bubbles1})
 \es{sigProp}{
\langle\sigma(x)\sigma(0)\rangle_\lambda=&\int \frac{d^d{p}}{(2\pi)^d}e^{ip\cdot x}\Big[D_\sigma(p)+D_\sigma(p)\mathbb{B}_{\phi^2}(p)D_\sigma(p) +\dots\Big]+O(N^{-1})\\
=&-\int \frac{d^d{p}}{(2\pi)^d}\frac{2\lambda e^{ip\cdot x}}{1-2\lambda \frac{
	|p|^{d-4}2^{1-2d}\pi^{\frac{3-d}{2}}}{\Gamma(\frac{d-1}{2})\sin(\frac{\pi d}{2})}}+O(N^{-1})\,.
}
In the IR CFT at $\lambda\to\infty$, this becomes the propagator of a dimension two operator:
\es{sigProp2}{
\langle\sigma(x)\sigma(0)\rangle_\infty&=\frac{2^{d+3} \sin \left(\frac{\pi  d}{2}\right) \Gamma (\frac{d-1}{2})}{\pi^{\frac32}  \Gamma
   \left(\frac{d}{2}-2\right)}\frac{1}{|x|^4}+O(N^{-1})\,.
}

For the gauge field $A_i$, we must fix a gauge in order to write down its propagator. It is convenient to choose a gauge such that the IR limit $e\to\infty$ can be taken immediately. As discussed in \cite{Chester:2016ref} for the similar fermionic QED$_3$ case, a simple family of gauges that has this property involves an average over different gauge-fixings, which is implemented by adding to the action the non-local gauge-fixing term
\es{GFaction}{
\mathcal{S}_{\zeta}[A]=\begin{cases}\frac{1}{1-\zeta} \frac{\Gamma \left(\frac{d}{2}\right)^2}{8\pi^d (d-1)(d-2)} \int d^dxd^dy\frac{\partial_i A^i(x)\partial_j A^j(y)}{|x-y|^{2d-4}}\qquad\qquad \,d\neq3\\
 \frac{1}{1-\zeta}\left(\frac{\kappa^2}{\pi^4}+\frac{1}{64\pi^2}\right) \int d^3xd^3y\frac{\partial_i A^i(x)\partial_j A^j(y)}{|x-y|^{2}}\qquad\qquad d=3\,,\\
\end{cases}
}
where $\kappa\equiv k/N$, and we chose a different coefficient for $d=3$ for later convenience. This arises by averaging over different $\partial_i A^i(x) = \omega(x)$ gauges with a non-local weight proportional to $(1-\zeta)^{-1} \int d^d x d^d y \omega(x) \omega(y) / |x-y|^{2d-4}$, where $\zeta\in\mathbb{R}$; the limit $\zeta\to1$ recovers the usual Landau gauge $\partial_i A^i=0$. Using this gauge-fixing term, along with the Maxwell term in \eqref{gaugeS}, and for $d=3$ the CS term in \eqref{CSaction}, we can find the momentum-space gauge field propagator
\es{Aprop0}{
D_{ij}^\text{Max}(p)=\begin{cases}\frac{e^2}{p^2}\left(\delta_{ij}-\frac{p_ip_j}{|p|^2}\right)+(1-\zeta)\frac{2^{d-2} \pi ^{d/2}\Gamma (d)}{ \Gamma \left(2-\frac{d}{2}\right) \Gamma
   \left(\frac{d}{2}\right)^2}\frac{p_ip_j}{|p|^d}\qquad\qquad\qquad\qquad\;\;\;\, d\neq3\\
\frac{1}{\frac{p^2}{e^4}+\frac{\kappa^2}{4\pi^2}}\left[\frac{1}{e^2}\left(\delta_{ij}-\frac{p_ip_j}{|p|^2}\right)-\frac{\kappa}{2\pi}\varepsilon_{ijl}\frac{p_l}{|p|^2}\right]+(1-\zeta)\frac{16\pi^2}{64\kappa^2+\pi^2}\frac{p_ip_j}{|p|^3}\qquad d=3\,.
\end{cases}
}

\begin{figure}[t]
\centering
   \includegraphics{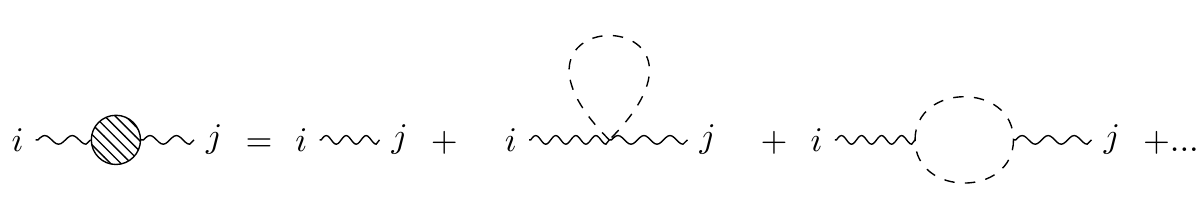}
   \caption{Bubble diagrams contributing to the effective photon propagator $\langle A_i(x)A_j(0)\rangle_{\zeta,e,k}$ at leading $O(N^0)$ order. Squiggly lines correspond to photon contractions $D^\text{Max}_{ij}$, and dashed lines to the scalars $\phi_I$.}
   \label{fig:bubbles2}
\end{figure}

The coupling $J_iA_i$ and the contact term $A_i^2\phi^2$ in \eqref{gaugeS} both contribute to the momentum space photon self-energy at one-loop order
\es{photonSE}{
\mathbb{B}_{ij}(p)&=\int \frac{d^dk}{(2\pi)^d}\left[-\frac{2\delta_{ij}}{|k|^2}+\frac{(2k+p)_i(2k+p)_j}{|k+p|^2|k|^2}\right]\\
&=-\frac{ \Gamma \left(2-\frac{d}{2}\right) \Gamma
   \left(\frac{d}{2}\right)^2}{2^{d-2} \pi ^{d/2}\Gamma (d)} |p|^{d-2} \left(\delta_{ij}-\frac{p_ip_j}{|p|^2}\right)\,,
}
where we see that the only role of the $A_i^2\phi^2$ term is to cancel a divergence in the contribution of the $J_iA_i$ term. As with $\sigma$, at leading order in $1/N$ we need to resum an infinite geometric series of bubble diagrams to find the effective photon propagator (see figure \ref{fig:bubbles2})
 \es{Aprop}{
&\langle A_i(x)A_j(0)\rangle_{\zeta,e,k}=  \int\frac{d^dp}{(2\pi)^d}  e^{ip\cdot x}\left[D_{ij}^\text{Max}(p)+D_{ik}^\text{Max}(p) \mathbb{B}_{kl}(p)D_{lj}^\text{Max}(p)+\dots\right]+O(N^{-1})\\
&= \begin{cases}\int\frac{d^dp}{(2\pi)^d}  e^{ip\cdot x}\Bigg[\frac{e^2  \left(\delta_{ij}-\frac{p_ip_j}{|p|^2}\right)}{|p|^2+\frac{ \Gamma \left(2-\frac{d}{2}\right) \Gamma
   \left(\frac{d}{2}\right)^2}{2^{d-2} \pi ^{d/2}\Gamma (d)} \frac{e^2}{|p|^{2-d}}}+\frac{2^{d-2} \pi ^{d/2}\Gamma (d)}{ \Gamma \left(2-\frac{d}{2}\right) \Gamma
   \left(\frac{d}{2}\right)^2}(1-\zeta)\frac{p_ip_j}{|p|^d}\Bigg]+O(N^{-1})\qquad d\neq3\\
\int\frac{d^3p}{(2\pi)^3}  \frac{e^{ip\cdot x}}{|p|}\Bigg[\frac{(\frac{1}{16}+\frac{|p|}{e^2})\left(\delta_{ij}-\frac{p_ip_j}{|p|^2}\right)-\frac{\kappa}{2\pi}\varepsilon_{ijl}\frac{p_l}{|p|}}{\left(\frac{1}{16}+\frac{|p|}{e^2}\right)^2+\frac{\kappa^2}{4\pi^2}}+\frac{16}{1+\frac{64\kappa^2}{\pi^2}}({1-\zeta})\frac{p_ip_j}{|p|^2}\Bigg]  +O(N^{-1})\quad\quad\;\; \,d=3\,,
\end{cases}
 }
 where the $k$ label is only there for $d=3$. In the IR CFT at $e\to\infty$ the expression simplifies and we obtain the momentum space propagator
  \es{Aprop2Mom}{
\langle A_i(p)A_j(-p)\rangle_{\zeta,\infty,k}&=\begin{cases} \frac{1 }{|p|^{d-2}} \frac{2^{d-2} \pi ^{d/2}\Gamma (d)}{ \Gamma \left(2-\frac{d}{2}\right) \Gamma
   \left(\frac{d}{2}\right)^2}\left(\delta_{ij}-\zeta\frac{p_ip_j}{|p|^2}\right)+O(N^{-1})\qquad\quad \,d\neq3\\
  \frac{1}{|p|}\frac{16\left(\delta_{ij}-\zeta\frac{p_ip_j}{|p|^2}\right)-\frac{128\kappa}{\pi}\varepsilon_{ijl}\frac{p_l}{|p|}}{1+\frac{64\kappa^2}{\pi^2}} +O(N^{-1})\qquad\qquad\;\; \,\;\;d=3
\end{cases}\\
 }
 from which we can go to position space to get 
 \es{Aprop2}{
\langle A_i(x)A_j(0)\rangle_{\zeta,\infty,k}&=\begin{cases}\frac{\Gamma (d)}{2 \Gamma \left(2-\frac{d}{2}\right) \Gamma \left(\frac{d}{2}\right)^3}\frac{(d-2-\zeta)\delta_{ij}+2\zeta\frac{x_ix_j}{|x|^2}}{|x|^2} +O(N^{-1})\qquad\qquad\qquad\; \;\;d\neq3\\
\frac{8}{\pi^2+{64\kappa^2}}\frac{1}{|x|^2}\left({(1-\zeta)\delta_{ij}+2\zeta\frac{x_ix_j}{|x|^2}}+\frac{\kappa}{4}\varepsilon_{ijl}\frac{x_l}{|x|}\right)  +O(N^{-1})\quad d=3\,.
\end{cases}
 }
 Note that the choice of gauge in \eqref{GFaction} allowed us to take $e\to\infty$ in the propagator, while other gauge choices such as $R_\xi$ gauge would give superficial divergences in this limit, which would only cancel after computing gauge-invariant observables.

Using these Feynman rules including the resummed propagators, we can compute correlation functions in scalar QED, in an expansion in $1/N$ similar to the large $N$ expansion of vector models.

\subsection{A subtlety in restricting to the singlet sector}
\label{restrict}

We will be interested in the $SU(N)$ singlet sector of scalar QED. In general $d$, we can simply restrict to this sector by hand, which has no nontrivial effects on $\mathbb{R}^d$. For $d=3$ and non-zero $k$, we need to be more careful. What we actually do is first restrict the theory of $N$ scalars to its $U(N)$-invariant sector, and then couple this sector to the $U(1)$ gauge field. We can perform the restriction in a rigorous way \cite{Giombi:2009wh} by coupling the theory to a $U(N)$ Chern-Simons term at infinite level before coupling to the new $U(1)$ gauge field, and we will use this to show that monopole operators cannot appear in the singlet sector (as we define it here) when $k\neq0$. Thus, for $d=3$ and non-zero $k$ our theory is not precisely the same as the $SU(N)$-singlet sector of scalar QED (which could have $SU(N)$-singlet monopole operators), but we will still call it by this name. In all other cases our theory includes the full $SU(N)$-singlet sector of scalar QED.

Let us start by considering the free theory with action $\mathcal{S}_\text{free}$ in $d=3$, with global symmetry $U(N)$. We can gauge this $U(N)$ by adding a $U(N)_{k'}$ Chern-Simons term for a gauge field $B_i$, with scalar couplings:
 \es{gaugeSB}{
  \mathcal{S}_{k'}&=\int d^3x\left[-\frac{ik'}{4\pi}\varepsilon_{ijl}(B_i{}^I_J\partial_jB_l{}^J_I-\frac{2i}{3}B_i{}_I^JB_j{}_J^KB_l{}_K^I)+J_i{}_I^JB_i{}_J^I+B_i{}_I^J B_i{}_J^I\phi_K^*\phi^K\right]\,\\
  &=\int d^3x\left[-\frac{iNk'}{4\pi}\varepsilon_{ijl}B_i\partial_jB_l+J_iB_i+NB_iB_i\phi^*_I\phi^I+\text{$SU(N)$ terms}\right]\,,\\
 }
 where $B_i{}^{I}_J$ is the $U(N)$ gauge field, $J_i{}_I^J\equiv i(\phi_I^*\partial_i\phi^J-\phi^J\partial_i\phi_I^*)$ is the $U(N)$ current, and in the second line we separated out the $SU(N)\subset U(N)$ terms from the $U(1)$ gauge field $B_i\equiv \frac{B_i{}_I^I}{N}$, which couples to the $U(1)$ current $J_i$ defined in \eqref{gaugeS}. The gauge-invariant operators are then all singlets of $U(N)$, as we want, but the theory is modified by the extra $B_i$ fields, and in particular it has new monopole operators charged under a topological $U(1)_{T'}$ global symmetry generated by the dual of the field strength of $B_i$. We can then take the $k'\to\infty$ limit, which on $\mathbb{R}^3$ removes all dynamical effects of the gauging, including the existence of monopoles charged under the $U(1)_{T'}$, and we obtain precisely the desired restriction to $U(N)$ singlets.
 
From the $U(N)$-singlet sector of the free theory, we can then get the theory we are interested in (the $SU(N)$-singlet sector of scalar QED at level $k$) by adding $\mathcal{S}_{e\to\infty}$, $\mathcal{S}_k$, and $\mathcal{S}_{\lambda\to\infty}$ as discussed before. For $k\neq0$, monopole operators (charged under $U(1)_T$) carry $k$ units of the gauge $U(1)$ charge, and so to be gauge-invariant under the $A$ gauging, monopoles needs to be dressed by composites of $\phi^I$ that will cancel their $U(1)$ charge. But then the dressed monopoles would also be charged under the original $B$ gauging due to the dressing. Thus, our theory includes no gauge-invariant monopoles for $k\neq0$. For $k=0$, the monopole operators of QED are uncharged under $U(1)$, so they are allowed. Similarly for $d\neq3$ where we have no Chern-Simons terms, the codimension-3 monopole operators discussed above are also allowed.

  \subsection{The bi-local formalism}
 \label{sec:bilocal_QED}
 
 We will now show how the singlet sector discussed above can be usefully described in terms of bi-local variables, which can then be naturally translated into the bulk in the later sections. We start by reviewing the bi-local formalism for the free theory with action $\mathcal{S}_\text{free}$ in \eqref{freeS} following \cite{Aharony:2020omh,Das:2003vw,Jevicki:1979mb,deMelloKoch:2018ivk}. All $U(N)$ invariants of $N$ free scalars can be written in terms of the bi-local field \eqref{defg}. We can then change variables in the path integral from $\phi^I$ to $G$ to get the partition function
\es{gaction}{
	Z = \int DG(x_1,x_2) \exp(-\mathcal{S}_\text{free}[G]-\mathcal{S}_\text{Jac}[G])\,,\qquad \mathcal{S}_\text{Jac}=-(N-V)\text{Tr}(\log(G))\,,
}
where we regularized the path integral by placing our field theory on a lattice of $V$ points, such that $G(x_1,x_2)$ is a Hermitian $V\times V$ matrix, and $\mathcal{S}_\text{Jac}$ is the non-trivial Jacobian (we drop all factors that do not depend on $G$). Note that \eqref{gaction} is only correct for $N\geq V$, which applies to the large $N$ expansion we consider in this paper, while for $N<V$ $G(x_1,x_2)$ must obey complicated non-linear constraints. The continuum limit is reached by taking $V\to\infty$, in which case matrix traces become continuum integrals as ${\rm Tr}(G) \equiv \int d^d x G(x,x)$ and $(GH)(x_1,x_2) \equiv \int d^d x_3 G(x_1,x_3) H(x_3,x_2)$. In particular, the continuum $\mathcal{S}_\text{free}[G]$ may be written as
\es{freeS2}{
 \mathcal{S}_\text{free}[G]=N\int d^dx_1 \partial_{1,i} \partial_{2,i}G(x_1,x_2)\vert_{x_2=x_1}\,.
} 
Since both $\mathcal{S}_\text{free}[G]$ and the Jacobian include terms proportional to $N$, we can perform a $1/N$ saddle point expansion by taking the large $N$ limit first, where all physical observables should be independent of the regularization parameter $V$. In particular, we expand $G$ around its large $N$ saddle point values given by the propagator $G_0(x_1,x_2)$ in \eqref{G0} as
\es{eta_def}{
	G(x_1,x_2)= G_{0}(x_1,x_2)+\frac{1}{\sqrt{N}}\eta(x_1,x_2)\,.
}
The free bi-local action \eqref{gaction} in terms of the fluctuation $\eta$ now gives (up to additive constants)
\es{freeEta}{
\mathcal{S}_\text{free}[\eta]={\sqrt{N}}\text{Tr}\left( G_{0}^{-1}\eta\right)\,,
}
and
\es{JacEta}{
 \mathcal{S}_\text{Jac}[\eta]&=-\left(N-V\right)\log\left(1+\frac{1}{\sqrt{N}} G_{0}^{-1}\eta\right)\\
 &=-\sum_{n=1}^{\infty}\frac{\left(-1\right)^{n+1}}{n}N^{1-\frac{n}{2}}\text{Tr}\left(\left( G_{0}^{-1}\eta\right)^{n}\right)-V\sum_{n=1}^{\infty}\frac{\left(-1\right)^{n}}{n}N^{-\frac{n}{2}}\text{Tr}\left(\left( G_{0}^{-1}\eta\right)^{n}\right)\,,
}
where we expanded in large $N$. Note that the first $n=1$ term cancels the linear $\mathcal{S}_\text{free}[\eta]$ as we expect from a saddle point solution. We can then write down Feynman rules where the propagator is given by the $n=2$ bare term in \eqref{JacEta}, and we have bare $n$-point vertices for $n\geq3$, as well as counterterm (multiplied by $V$) $n$-point vertices for $n\geq1$. As shown in \cite{Aharony:2020omh}, these Feynman rules lead to the expected correlation functions for the free theory.

We can extend the bi-local formalism to QED by simply writing all $\phi$-dependent terms in the actions we wrote above using $G$ as
\es{actionG}{
\mathcal{S}_e[G,A]&=\int d^dx\Bigg[\frac{1}{4e^2}F_{ij}^2(x)+J_i[G]A_i(x)+A_i^2(x)G(x,x)\Bigg] \,,\\
 \mathcal{S}_{\lambda}[G,\sigma]&=\int d^dx\Bigg[-\frac{1}{4\lambda}\sigma^2(x)+\frac{\sqrt{N}}{2}\sigma(x)G(x,x)\Bigg]\,,\\
}
where we write the $U(1)$ current in terms of $G$ as $J_i[G]\equiv i{\sqrt{N}}(\partial_{2,i}-\partial_{1,i})G(x_1,x_2)\vert_{x_2=x_1}$. Note that $\mathcal{S}_\text{Jac}[G]$ is gauge-invariant by itself, and then gauge-invariance of the other terms works exactly the same as in the $\phi$ language. We would like to now expand $G$ around $G_0$ in \eqref{G0} (which is not gauge-invariant, but is still a saddle point in the gauge-fixing that we perform as described above) to get
\es{actionEta}{
\mathcal{S}_e[\eta,A]&=\int d^dx\Bigg[\frac{1}{4e^2}F_{ij}^2(x)+J_i[\eta]A_i(x)+J_i[G_0]A_i(x)+\frac{{A_i^2}(x)\eta(x,x)}{\sqrt{N}}+A_i^2(x)G_0(x,x)\Bigg] \,,\\
 \mathcal{S}_{\lambda}[\eta,\sigma]&=\int d^dx\Bigg[-\frac{1}{4\lambda}\sigma^2(x)+\frac{1}{2}\sigma(x)\eta(x,x)+\frac{\sqrt{N}}{2}\sigma(x)G_0(x,x)\Bigg]\,,\\
}
where $J_i[\eta]\equiv i(\partial_{2,i}-\partial_{1,i})\eta(x_1,x_2)\vert_{x_2=x_1}$. This includes divergent terms involving $G_0(x,x)$ and its derivatives, which are not regularized by the lattice regulator used so far. For $S_\lambda$, we can cancel $\sigma(x)G_0(x,x)$ at each order with a linear term in $\sigma$ (or, equivalently, with a mass counterterm). For $S_e$, we expect that $J_i[G_0]=0$ in any Lorentz-invariant regularization\footnote{A gauge-invariant way of cancelling this term is to couple $N$ new scalars $\tilde\phi_I$ to $A_i$ with opposite charge as $\phi_I$ and large mass $m$. We can then express $\tilde\phi_I$ in terms of bi-locals $\tilde G$ in the usual way except the saddle point is now given by the massive free propagator
\es{newG0}{
\tilde G_0(x,0)=\frac{\Gamma(d/2-1)}{4\pi^{d/2}}\frac{e^{-m |x|}}{|x|^{d-2}}\,,
}
and the $J_iA_i$ terms in the $\tilde G$ version of \eqref{actionEta} come with opposite sign. If we now take the limit $m\to\infty$, then we see that $\tilde G_0(x_1,x_2)\to0$ except where $x_1\to x_2$ such that $\tilde G_0\to G_0$, so the only effect of these new scalars is to cancel the $J_i[G_0]A_i$ term.}. In the following, we will simply drop the divergent terms that can be cancelled in these ways, and thus define
\es{actionEta2}{
\mathcal{S}_e[\eta,A]&\equiv\int d^dx\Bigg[\frac{1}{4e^2}F_{ij}^2(x)+J_i[\eta]A_i(x)+\frac{{A_i^2}(x)\eta(x,x)}{\sqrt{N}}+A_i^2(x)G_0(x,x)\Bigg] \,,\\
 \mathcal{S}_{\lambda}[\eta,\sigma]&\equiv\int d^dx\Bigg[-\frac{1}{4\lambda}\sigma^2(x)+\frac{1}{2}\sigma(x)\eta(x,x)\Bigg]\,.\\
}
Note that we have kept the counterterm $A_i^2(x)G_0(x,x)$ in the action, because it is needed to cancel the divergence in the contribution of $J_i[\eta]A_i$ to the photon self energy. In particular, $A_i^2(x)G_0(x,x)$ contributes the first divergent term in \eqref{photonSE}, while $J_i[\eta]A$ contributes the second term. The $\frac{{A_i^2}(x)\eta(x,x)}{\sqrt{N}}$ term does not contribute at leading order in $1/N$, but will contribute at sub-leading orders. In fact, for the $\mathbb{CP}^{N-1}$ model \eqref{CPNS} we can make the field redefinition (after gauge-fixing) $\sigma(x)\mapsto \sigma(x) - \frac{2 A_i^2(x)}{\sqrt{N}}$ which cancels $\frac{{A_i^2}(x)\eta(x,x)}{\sqrt{N}}$ entirely (for $\lambda \to \infty$). In this way the Feynman diagrams of the theory can be packed in terms of the effective $\sigma$ and $A$ propagators in \eqref{sigProp2} and \eqref{Aprop}, respectively.

Our action is written in a specific gauge choice discussed above, using the non-gauge-invariant bi-local variable $\eta(x_1,x_2)$. In order to construct gauge-invariant variables, we need to look at the limit $x_2 \to x_1$ of $\eta(x_1,x_2)$ to obtain gauge-invariant local operators, or alternatively to dress $G(x_1,x_2)$ with a Wilson line between $x_1$ and $x_2$. We can also compute correlation functions of $\eta(x_1,x_2)$'s using our gauge-fixed action, but generally these are not meaningful, because we are averaging over different gauge choices in which $G(x_1,x_2)$ corresponds to different operators\footnote{In the special case $\zeta=1$ where we just have Landau gauge, these correlation functions are meaningful, but they correspond to some complicated non-local gauge-invariant operators.}.

\section{The AdS/CFT map} 
\label{sec:map}

In this section, we discuss the exact AdS/CFT map that we will use in the next section to write down the bulk dual of the $SU(N)$-singlet sector of scalar QED. We start by reviewing the off-shell AdS/CFT map derived in \cite{Aharony:2020omh}, which naturally acts on the CFT bi-local $\eta(x_1,x_2)$. We will then show how this map implies that local single-trace operators of any spin $J$ in any dimension $d$ also map off-shell to the bulk in a simple way, generalizing the $d<4$ and $J=0$ case proven in \cite{Aharony:2020omh}. For $J=1$, this off-shell map will then be used to write the bulk dual of scalar QED in the next section.

\subsection{Review of the bi-local map}
\label{sec:review}

We begin by briefly reviewing the AdS/CFT map of the free scalar theory \cite{Aharony:2020omh}. The map is given by expanding the bi-local fluctuation $\eta( x_1 , x_2)$ on one side, and the spin $J$ transverse traceless AdS fields $\Phi_J(x,z)$ on the other side, in terms of the same irreducible representations (labeled by $\Delta,J,y$) of the conformal group, so that the exact map between $\eta( x_1 , x_2)$ to $\Phi_J(x,z)$ is given by the convolution of the basis elements in each space.\footnote{For later convenience, we will not use the embedding space formalism used in \cite{Aharony:2020omh}.} 

\begin{figure}[!ht]
\centering
	\includegraphics{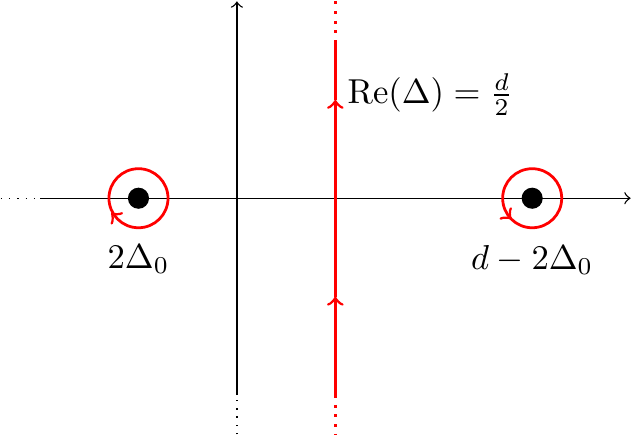}
	\caption{The deformation of the contour $\gamma_0$ from the principal series $\text{Re}(\Delta)=\frac{d}{2}$ to include $\Delta=2\Delta_0$ and exclude $d-2\Delta_0$, when $2\Delta_0<d/2$. In general we should deform the contour $\gamma_J$ for any $2\Delta_0+2n+J<\frac{d}{2}$ and $n=0,1,\cdots$.}
	\label{fg:Delta_contour}
\end{figure}

We expand $\eta( x_1 , x_2)$ in the complete basis
\es{eta_decomp}{
	\eta(x_1,x_2) = \sum_{J=0}^\infty \int_{\gamma_J}\frac{d\Delta}{2\pi i} \int d^d y  \, C_{\Delta,J}\left(y\right)
	\thpnorm{x_1}{x_2}{\Delta,J}{y}\,,
}
where the basis elements are ``3-point functions''\footnote{Note that while the harmonic basis resembles a three point function, it does not correspond to a correlator in a physical CFT, and is simply a useful basis for conformally-covariant functions.} of scalar operators $\mathcal{O}_{\Delta_0}$ and $\hat{\mathcal{O}}_{\Delta_0}$ that have the same scaling dimension $\Delta_0=\frac{d-2}{2}$ as a free scalar field, namely
\es{3point}{
	\thpnormind{x_1}{x_2}{\Delta,J}{x_3}{i_1,...,i_J} = \frac
	{Z^{i_1}...Z^{i_J}-\text{traces}}
	{x_{12}^{2\Delta_{0}-\Delta+J}x_{13}^{\Delta-J}x_{23}^{\Delta-J}}\,,
}
where $Z^{i} \equiv \frac{x_{13}^i}{x_{13}^2}-\frac{x_{23}^i}{x_{23}^2}$, we define $x_{12}\equiv |x_1-x_2|$, and we will in general suppress spin indices for simplicity. The contours $\gamma_J$ of the $\Delta$ integrals in \eqref{eta_decomp} go over the principal series $\Delta=\frac d2+is$ for real $s$, except for $J=0$ and $d<4$ where we deform the contour as shown in Figure \ref{fg:Delta_contour} to ensure that the pole $2\Delta_0=d-2$ of the scalar single-trace operator in the free theory appears on the same side of the contour as the other spin $J$ single-trace operators with scaling dimension $d-2+J$. This expansion exists when $\eta(x_1,x_2)$ satisfies the conditions
\begin{enumerate}
	\item $\lim_{x_2\to x_1}\eta(x_1,x_2)$ should be finite.
	\item At large $|x_1+x_2|$ (and fixed difference) $\eta(x_1,x_2)$ should decay.\footnote{In \cite{Aharony:2020omh}, a related map was also derived for $\eta(x_1,x_2)$ that do not satisfy this condition.}
	\item At large $|x_1|$ (or $|x_2|$) and fixed $x_2$ ($x_1$)
	\begin{equation}
		\eta(x_1,x_2) \sim \left|x_{1}\right|^{-2\Delta_{0}}\cdot \text{Power series in }\ensuremath{\frac{1}{|x_1|}}\,. \label{eq:bc_eta}
	\end{equation}
	\item $\eta(x_1,x_2)$ must be smooth.
\end{enumerate}

For $d>4$,\footnote{We expect that $d=4$ should be an analytic continuation of our results for $d>4$, as is generally the case in harmonic analysis \cite{Caron-Huot:2017vep}, but we will not discuss this case in detail.} we can use the orthogonality and completeness relations of the 3-point basis, as reviewed in Appendix \ref{3pointBasis}, to invert \eqref{eta_decomp} and write $C_{\Delta,J}(y)$ in terms of $\eta(x_1,x_2)$ as
\es{coeff}{
d>4:\qquad	C_{\Delta,J}(y)=\frac12\frac{1}{N_{\Delta,J}}\int d^dx_{1}d^dx_{2}\, \eta\left(x_{1},x_{2}\right)
	\thptild{x_1}{x_2}{\tilde\Delta,J}{y}\,,
}
where the normalization $N_{\Delta,J}$ is given in \eqref{CFTNorm}, and tildes over dimensions denote $\tilde\Delta=d-\Delta$. The shadow relation \eqref{eq:shadow_trans} implies that only half of the coefficients $C_{\Delta,J}(y)$ are independent along the contour $\gamma_J$. We define the physical $C_{\Delta,J}(y)$ to have $\text{Im}(\Delta)\geq0$, and then the shadow $C_{\tilde\Delta,J}(y)$ are related to them by
\es{basExtend}{
	C_{\tilde \Delta,J}(y) = \frac{1}{S^{(\tilde\Delta,J)}_{\Delta_0,\Delta_0} } 
	\int d^d y^\prime 
	\tp{\Delta,J}{y}{y^\prime}
	C_{\Delta,J}(y^\prime)\,, 
}
where the shadow coefficient $S^{(\tilde\Delta,J)}_{\Delta_0,\Delta_0}$ is given in \eqref{S}, and the ``2-point function'' is defined as
\es{2point22}{
\langle\mathcal{O}_{\Delta,J}^{i_1\dots i_J}(x_1)\mathcal{O}_{\Delta,J}^{i'_1\dots i'_J}(x_2)\rangle&=\frac{I^{i_1i'_1}(x_{12})\dotsb I^{i_Ji'_J}(x_{12})}{x_{12}^{2\Delta}}-\text{traces}\,,\\
}
where $I^{ii'}\equiv\delta^{ii'}-2\frac{x^{i} x^{i'}}{|x|^2}$. Hermiticity of $\eta(x_1,x_2)$ then implies that $C^*_{\Delta,J}(y)=(-1)^JC_{\tilde\Delta,J}(y)$. 

For $d<4$, the integral in \eqref{coeff} does not converge. As explained in \cite{Aharony:2020omh}, we can avoid this divergence by considering the auxiliary bi-local
\es{aux}{
		\tilde\eta(x_1,x_2) =\left( \frac{\Gamma(\frac{d-2}{2})}{4\pi^{\frac{d}{2}}}\right)^2 \nabla^2_1\nabla^2_2 \eta(x_1,x_2)\,,
}
which can be expanded in the harmonic basis as
\es{eta2_decomp}{
\tilde\eta(x_1,x_2) = \sum_{J=0}^\infty \int_{P.S.}\frac{d\Delta}{2\pi i} \int d^d y  \, \tilde C_{\Delta,J}(y)
	\thptild{x_1}{x_2}{\Delta,J}{y}\,,
	}
	where the contour is now the principal series for all $J$, and 
	\es{coeff2}{
d<4:\qquad	\tilde C_{\Delta,J}\left(y\right)=\frac12\frac{1}{N_{\Delta,J}}\int d^dx_{1}d^dx_{2}\, \tilde\eta\left(x_{1},x_{2}\right)
	\thpnorm{x_1}{x_2}{\tilde\Delta,J}{y}
	}
	is now convergent for $d<4$, unlike \eqref{coeff}. The original expansion of $\eta(x_1,x_2)$ in \eqref{eta_decomp} still holds for $d<4$ provided that we identify
\es{Ctilde}{
	d<4:\qquad C_{\Delta,J}(y) \equiv \frac{16\pi^d \Gamma^2(\frac{d}{2}-\Delta_0)}{\Gamma^2(\Delta_0)\lambda_{\Delta,J}} \,
	\tilde C_{\Delta,J}(y)\,,
}
where $\lambda_{\Delta,J}$ is the eigenvalue of the bi-local Laplacian in the conformal basis and is given for any $d$ by
\es{lam}{
\lambda_{\Delta,J}=\left(M_{\Delta,J}^{2}-M_{d+J,J}^{2}\right)\left(M_{\Delta,J}^{2}-M_{d+J-2,J}^{2}\right)\,,\qquad M_{\Delta,J}^{2}\equiv \Delta\left(\Delta-d\right)-J\,.
}

In AdS space we use Poincar\'e coordinates ($ds^2 = (dx^i dx^i + dz^2) / z^2$), and in our formalism the metric in the bulk is fixed, with traceless transverse spin $J$ fields $\Phi_J(x,z)$ (including a spin-two graviton) propagating on this fixed background.
We define the mapping from the CFT to the bulk by expanding the bulk fields $\Phi_J(x,z)$ in the complete basis
\es{bulk_comp}{
\Phi_{J}(x,z)=\int_{\gamma_J}\frac{d\Delta}{2\pi i}\int 
	d^dy
	f_{\Delta,J}C_{\Delta,J}(y)G_{\Delta,J}(x,z|y)\,,
	}
	where we identified the bulk coefficient $C_{\Delta,J}(y)$ with the same coefficient appearing in the bi-local expansion, up to a multiplicative factor $f_{\Delta,J}$ that is not fixed by conformal symmetry. This identification implies that the contour $\gamma_J$ in \eqref{bulk_comp} is the same as the one in the bi-local expansion. The basis elements in \eqref{bulk_comp} are the bulk-to-boundary propagators in AdS space, defined by the differential equation
\es{eq:bulk_boundary_DE}{
	&(\nabla^2_{x,z} -M^2_{\Delta,J})G _{\Delta,J}(x,z|y)=  0\,,\qquad \nabla^2_{x,z} \equiv z^{d+1}\partial_{z}\left(z^{-d+1}\partial_{z}\right)+z^{2}\nabla_x^2\,,\\
}
and by the $z\to0$ boundary condition\footnote{When $\Delta=\tilde\Delta$, as can happen for the scalar bulk field in $d=4$, the second term will have a $\log(z)$, which distinguishes its scaling in $z$ from the first term.}
 \es{eq:2_point_limit}{
  	G_{\Delta,J}(x,z|y)=z^{\Delta-J} 
 	\langle\mathcal{O}_{\Delta,J}(x)\mathcal{O}_{\Delta,J}(y)\rangle + z^{d-\Delta-J} S_B ^{\Delta,J} \delta_J^{TT}(x-y)+ \dots\,,
}
where $\delta_J^{TT}(x)$ denotes a delta function with $2J$ suppressed lower indices for spin $J$ traceless transverse functions on the boundary $\mathbb{R}^d$,\footnote{For instance, when $J=1$ we can define $\delta^{TT}_{1,ij}(x)\equiv\int \frac{d^dp}{(2\pi)^d} \,e^{ipx}(\delta_{ij}-\frac{p_ip_j}{|p|^2})$.} and we define the bulk shadow coefficient
\es{eq:bulk_shadow_coeff}{
	S_B^{ \Delta,J} 	\equiv \frac{\pi^{\frac{d}{2}} \Gamma( \Delta-\frac{d}{2})}
	{(J+ \Delta-1)\Gamma(\Delta-1)}\,.
}
The solution to \eqref{eq:bulk_boundary_DE} and \eqref{eq:2_point_limit} is by construction transverse
\es{eq:bulk_boundary_lorentz}{
\text{for $a=1,\cdots, J$}:\qquad	\nabla^{\mu_a}G^{i_1\cdots i_J}_{\Delta,J}(x,z|y)_{\mu_1\cdots \mu_a\cdots \mu_J} = 0\,, \qquad \nabla^\mu\equiv z^{d+3}\partial^\mu z^{-d-1} \,,
}
where $\mu=z,1,\cdots,d$ denotes bulk spin indices that we in general suppress for simplicity (we denote the index in the radial direction by $z$, this should not be confused with the value of the radial coordinate that we also denote by $z$).
See Appendix \ref{bulkBasis} for an explicit expression for $G_{\Delta,J}$, as well as the orthogonality, completeness, and shadow relations. 

The CFT shadow relation \eqref{eq:shadow_trans} and the bulk shadow relations \eqref{eq:shadow_trans_bulk} imply that $f_{\Delta,J}$ must satisfy the consistency condition
\es{bulk_norm_shadow_cond}{
		\frac{f_{\Delta,J}}{f_{\tilde \Delta,J} } = \frac{S_B^{\tilde \Delta,J}}{S^{(\tilde\Delta,J)}_{\Delta_0,\Delta_0}} =
		\frac{ \Gamma\left(\Delta+J\right)\Gamma^2\left(\frac{\tilde\Delta+J}{2}\right)}
	 	{ \Gamma\left(\tilde\Delta+J\right)\Gamma^2\left(\frac{\Delta+J}{2}\right)}\,,
}
where the bulk shadow coefficient is given in \eqref{eq:bulk_shadow_coeff}. The expansion \eqref{bulk_comp} assumes that $\Phi_J(x,z)$ decays in the small $z$ limit as $z^{\frac d2-J}$ in general. For $J=0$ and $d<4$ recall that the contour $\gamma_0$ allows the integrand to have an extra contribution at $\Delta=d-2$, in which case $\Phi_0(x,z)$ could decay as $z^{d-2}$. Finally, $\Phi_J(x,z)$ should decay at large $x$, which corresponds to $\eta(x_1,x_2)$ decaying at large $|x_1+x_2|$.\footnote{A map for non-decaying $\Phi_0$ was also derived in \cite{Aharony:2020omh}.} 

For $J>0$ or $d>4$, we can use the orthogonality and completeness of the propagators to write $C_{\Delta,J}(y)$ in terms of $\Phi_J(x,z)$ as
\es{coeffBulk}{
J>0\quad\text{or}\quad d>4:\qquad	C_{\Delta,J}(y)=\frac{1}{\alpha_Jf_{\Delta,J}N_{\Delta,J}}\int \frac{d^dx dz}{z^{d+1}} \Phi_{J}(x,z|y)G_{\tilde\Delta,J}(x,z|y)	\,,
}
where the normalization $\alpha_J$ is given in \eqref{eq:alpha_J_def}. This integral is not convergent for $J=0$ and $d<4$, but as shown in \cite{Aharony:2020omh} it can be replaced by the modified relation
\es{auxB}{
J=0\ \text{and}\ d<4:\quad C_{\Delta,0}(y)=\int\frac{d^dx dz}{z^{d+1}}\frac{G_{\tilde\Delta,0}(x,z|y)}{\lambda_{\Delta,0}\alpha_0f_{\Delta,0}N_{\Delta,0}}(\nabla^2_{x,z}-M_{d-2,0}^2)(\nabla^2_{x,z}-M_{d,0}^2)\Phi_0(x,z)\,,
}
where the AdS Laplacian is defined in \eqref{eq:bulk_boundary_DE}.

The CFT-to-AdS map now simply follows from a convolution of the bulk and CFT bases, and it takes the explicit form
\es{CFTtoAdS}{
d>4:\qquad	\Phi_{J}(x,z)& =\frac12
	\int_{P.S.}\frac{d\Delta}{2\pi i}\frac{f_{\Delta,J}}{N_{\Delta,J}}\int d^dy\int d^dx_1 d^dx_2  \, G_{\Delta,J}(x,z|y)
	\\
	&\qquad\qquad\times \thptild{x_1}{x_2}{\tilde\Delta,J}{y}\eta(x_1,x_2)\,,\\
d<4:\qquad	\Phi_{J}(x,z)& = \frac12 \int_{\gamma_J} \frac{d\Delta}{2\pi i}  \int d^dy \int d^dx_1 d^dx_2 \frac{f_{\Delta,J}}{\lambda_{\Delta,J}N_{\Delta,J}} \\
		&\quad \times  
		G_{\Delta,J}(x,z|y)
		\thpnorm{x_1}{x_2}{\tilde\Delta,J}{y} \, \nabla^2_1\nabla^2_2\eta\left(x_1,x_2\right)\,,
}
where for $d>4$ we plugged \eqref{coeff} into \eqref{bulk_comp}, while for $d<4$ we plugged \eqref{coeff2} and \eqref{Ctilde} into \eqref{bulk_comp}. The AdS-to-CFT map similarly comes from plugging \eqref{coeffBulk} into \eqref{eta_decomp}, and it takes the form
\es{AdStoCFT}{
\eta(x_1,x_2) & = 
			\sum_{J=0}^\infty 
			\int_{P.S.}\frac{d\Delta}{2\pi i} 
			\int d^dy
			\int \frac{d^dxdz}{z^{d+1}} 
			\frac{G_{\tilde{\Delta},J}(x,z|y)}{\alpha_J \, N_{\Delta,J} f_{\Delta,J}}
			\thpnorm{x_1}{x_2}{\Delta,J}{y}
			\Phi_{J}(x,z)\,,
}
where for $d<4$ and $J=0$ we need to replace the $J=0$ term by
\es{J0}{
			\int_{\gamma_0}\frac{d\Delta}{2\pi i} 
			\int d^dy \int \frac{d^dxdz}{z^{d+1}}  
			\thpnorm{x_1}{x_2}{\Delta,0}{y}
			 \frac{G_{\tilde{\Delta},0}\left(x,z|y\right)}{\alpha_0 N_{\Delta,0} \lambda_{\Delta,0} f_{\Delta,0}}\\
			\times 
			\left(\nabla_{x,z}^{2}-M_{d-2,0}^{2}\right)\left(\nabla_{x,z}^{2}-M_{d,0}^{2}\right)\Phi_{0}(x,z)\,,
}
due to the modified expansion in \eqref{auxB}.

\subsection{Off-shell map of single-trace operators }
\label{sec:newmap}

We will now discuss how the off-shell bi-local map reviewed above can be used to derive a simple off-shell map for singlet local operators in the CFT. We start by reviewing the derivation of the map for $J=0$ and $d<4$ that was shown in \cite{Aharony:2020omh}, and then discuss how to generalize this to general $J$ and $d>2$.

Spin $J$ single-trace singlet local operators $S^J_{i_1\cdots i_J}(x)$ in the free theory are defined in terms of the bi-local as 
\es{st}{
S_{i_1\cdots i_J}^J(x_1)\equiv \lim_{\eps\to 0}D_{i_1\cdots i_J}^{J,(x_1,x_2)}\eta(x_1,x_2)\vert_{x_2=x_1+\eps\hat e}\,,
}
where $\hat e$ is an arbitary unit vector, and the bi-local differential operator $D_{i_1\dots i_J}^{J,(x_1,x_2)}$ can be found in \cite{Craigie:1983fb} and is fixed such that $S_J(x)$ is a conformal primary normalized with two-point function
\es{2point2}{
\langle S_{i_1\dots i_J}^J(x_1)S^J_{j_1\dots j_J}(x_2)\rangle=&\mathfrak{a}_{J}\left(\frac{I_{j_1(i_1}(x_{12})\dotsb I_{i_J)j_J}(x_{12})}{x_{12}^{2(d-2+J)}}-\text{traces}\right)\,,\\
 I_{ij}\equiv\delta_{ij}-2\frac{x_i x_j}{x^2}\,,\quad \mathfrak{a}_{J}=&\frac{\pi ^{\frac{1}{2}-d}  \Gamma \left(\frac{d}{2}+J-1\right) \Gamma
   (d+J-3)}{2^{d+J}\Gamma (J+1) \Gamma \left(\frac{d-3}{2}+J\right)}\,.
}
For instance, for $J=0,1$ we have 
\es{J01}{
D^{0,(x_1,x_2)}=1\,,\qquad D^{1,(x_1,x_2)}_i=\frac{1}{2\sqrt{d-2}}(\partial_{i,x_2}-\partial_{i,x_1})\,,
}
such that for $J=0$ we recover the coefficient $\mathfrak{a}_{0}=\frac{\Gamma(d/2-1)^2}{16\pi^d}$ in the scalar 2-point function computed from \eqref{G0}, while for $J=1$ we identify $S^1_i(x)=-\frac{i}{2\sqrt{d-2}}J_i[\eta]$ with $J_i[\eta]$ defined below \eqref{actionEta}.

As shown in \cite{Aharony:2020omh}, we can use the expansions \eqref{eta_decomp} and \eqref{bulk_comp} to show that 
\es{Sdecomp}{
	S^J_{i_1\dots i_J}(x_1) =& \lim_{\eps\to0}\Big[\int_{\gamma_{J}}\frac{d\Delta}{2\pi i} \frac{2S^{(\tilde\Delta,J)}_{\Delta_0,\Delta_0}C_{i_1\dots i_{J}}^{\tilde\Delta,J}(x_1)}{\eps^{2\Delta_0+J-\Delta}}+ \\ & \qquad\qquad \sum_{J'=0}^\infty \int_{\gamma_{J'}}\frac{d\Delta}{2\pi i}\int d^dy C_{j_1\dots j_{J'}}^{\Delta,J'}(y)\frac{(\text{$\hat e$-dependent})_{i_1\dots i_J;j_i\dots j_{J'}}}{\eps^{2\Delta_0+J'-\Delta}}\Big]\,,\\
	\Phi_{J,i_1\dots i_J}(x_1,\eps)=&2\int_{\gamma_J} \frac{d\Delta}{2\pi i}   f_{\Delta,J}\eps^{\Delta-J} \left[S_{\Delta_0,\Delta_0}^{(\tilde\Delta,J)}C_{i_1,\dots i_J}^{\tilde\Delta,J}(x_1)+O(\eps)\right]\,.\\
}
In general, it is difficult to perform the $\Delta$ integrals in \eqref{Sdecomp}, since we know very little about general $C_{\Delta,J}$. For instance, the contour $\gamma_{J'}$ for $d>4$ or for $J'>0$ and $d\leq4$ is the principal series $\text{Re}(\Delta)=d/2$, so along this contour the leading term $\eps^{2-d/2-J'}$ diverges for the free theory. To get the finite answer for $S^J_{i_1\dots i_J}(x_1) $ that we expect, there must be complicated cancellations. For $d<4$ and $J'=0$, however, recall that the contour $\gamma_0$ includes a deformation from the principal series to include the pole $\Delta=d-2<d/2$. Since the principal series contribution goes to zero as $\eps^{2-d/2}$ in this case, we know even off-shell that the only contribution as $\eps \to 0$ to the integrals in \eqref{Sdecomp} comes from the $\Delta=d-2$ pole, which is the only pole on the other side of the principal series. This yields the off-shell relation 
\es{toshow2}{
 S^0(x)\equiv\eta(x,x)=\frac{1}{f_{d-2,0}}\lim_{\eps\to0}\eps^{2-d}\Phi_0(x,\varepsilon)\,,
}
which in particular continues to hold under deformations of the theory. 

We can generalize this relation to general $d>2$ and $J$ using a different argument. Consider deforming the bi-local CFT action by $S^J$ coupled to a source $A^J$ as
\es{deform}{
\mathcal{S}_{A_J}=\int d^dx \,A^J_{i_i\dots i_J}S^J_{i_1\dots i_J}\,.
}
We can compute the VEV of $\eta(x_1,x_2)$ under this deformation (in the large $N$ limit) as
\es{vev}{
\langle\eta(x_1,x_2)\rangle_{A_J}&=\int d^dy \,A^J_{i_i\dots i_J}(y)D_{i_1\cdots i_J}^{J,(x_1,x_2)}G_0(x_1,y)G_0(y,x_2)\\
&=\mathfrak{a}_J\int d^dy \,A^J_{i_i\dots i_J}(y) \thpnormind{x_1}{x_2}{d-2+J,J}{y}{i_1,...,i_J}\,,
}
where the second equality follows from the definitions \eqref{G0}, \eqref{3point}, and \eqref{st}. As $x_2\to x_1$, this VEV diverges as $S^{(d-2+J,J)}_{\Delta_0,\Delta_0}A^J(x_1)/x_{12}^{d-4+2J}$. At leading $N\to\infty$, this implies the off-shell singularity
\es{sing}{
\lim_{ \varepsilon\to0} \varepsilon^{4-d-2J} \eta(x_1,x_2)\vert_{x_2=x_1+\varepsilon\hat e}=S^{(d-2+J,J)}_{\Delta_0,\Delta_0}\mathfrak{a}_J{A^J_{i_i\dots i_J}(x_1)\hat e_{i_1}\dots \hat e_{i_J}}{}\,,
}
which is singular for all $d>2$ and $J$ except $J=0$ for $d<4$. In the original path integral in terms of the local field $\phi_I(x)$, each $\phi_I(x)$ does not couple to the others, which implies that the off-shell behavior of $\phi_I(x)$, and thus $\eta(x_1,x_2)$, is independent of $N$, so \eqref{sing} in fact holds for finite $N$. Recall that the conditions \eqref{eq:bc_eta} to expand $\eta(x_1,x_2)$ in terms of $C_{\Delta,J}(y)$ require that $\eta(x_1,x_2)$ be finite as $x_2\to x_1$, so we must modify our AdS/CFT map in the presence of this source for all $d>2$ and $J$ except $J=0$ for $d<4$. We can cancel the divergence \eqref{sing} by modifying the contour $\gamma_J$ in \eqref{eta_decomp} to include a piece around $\Delta=d-2+J$ with
\es{modC}{
C_{d-2+J,J}(y)\vert_{A_J}=\mathfrak{a}_JA^J(y)\,,
}
which will cancel the divergent VEV in \eqref{vev}. In the bulk, this modification of $\gamma_J$ will give the VEV
\es{modBulk}{
\langle\Phi_J(x,z)\rangle\vert_{A_J}=\mathfrak{a}_J f_{d-2+J,J} \int d^dy\, A^J(y)G_{d-2+J,J}(x,z|y)\,,
}
which follows from modifying the contour in \eqref{bulk_comp}, and is also what we would get by naively mapping \eqref{vev} using the un-modified map \eqref{CFTtoAdS} and the CFT orthogonality relation \eqref{ortho}. We can then take $z\to0$ to find the modified off-shell bulk boundary condition
\es{bulkBound}{
\lim_{\varepsilon\to0}\varepsilon^{2J-2}\Phi_J(x,\varepsilon)=\mathfrak{a}_Jf_{d-2+J,J} S_B^{d-2+J,J}A^J(x)\,,
}
which follows from \eqref{eq:2_point_limit}. We could equivalently use the standard bulk boundary conditions and instead add to the bulk action the source term
\es{deformBulk}{
\mathcal{S}^{bulk}_{A_J}=\int \frac{d^dxdz}{z^{d+1}} \,A^J_{i_i\dots i_J}(x) \frac{1}{f_{d-2+J,J}}\lim_{\varepsilon\to0}\varepsilon^{2-d}\Phi_J(x,\varepsilon)\,.
}
Now, comparing \eqref{deformBulk} to \eqref{deform} for general $A^J(x)$ implies the off-shell relation
\es{toshow}{
 S^J(x)=\frac{1}{f_{d-2+J,J}}\lim_{\eps\to0}\eps^{2-d}\Phi_J(x,\varepsilon)\,,
}
which for $J=0$ for $d<4$ was what we previously showed in \eqref{toshow2}, and for all other $d>2$ and $J$ follows from the modified boundary condition argument. This off-shell map generalizes the on-shell relation previously shown for general $d>2$ and $J$ in \cite{Aharony:2020omh}.

\section{The bulk dual of scalar QED} 
\label{sec:QED_bulk}

We will now use the AdS/CFT map of the previous section to write the action for the bulk dual of scalar QED. We will first review the bulk action for the free and critical $U(N)$ theories, which were derived in \cite{Aharony:2020omh}. We will then use the off-shell map of single-trace local operators for $J=0,1$ to show that the bulk QED action is given by a simple deformation of the bulk dual of the free theory. Finally, we will compute correlation functions in the bulk and show that the bulk duals of the free theory and QED only differ by the boundary conditions of the bulk $J=0,1$ two-point function, as anticipated in \cite{Klebanov:1999tb,Witten:2001ua,Witten:2003ya,Chang:2012kt,Berkooz:2002ug}, where the $J=1$ two-point function is sensitive to the gauge-fixing in the CFT.

\subsection{The bulk action}
\label{sec:action}

We start by reviewing the bulk action for the free theory, which in the bi-local language had an infinite number of terms given by \eqref{freeEta} and \eqref{JacEta}. The bi-local AdS/CFT map translates each of these terms to the bulk, where for general $f_{\Delta,J}$ they take a complicated non-local form given explicitly in \cite{Aharony:2020omh}. For the special choice 
\es{flocal}{
f_{\Delta,J}^\text{local} = \left( (-1)^J \frac{S^{\tilde\Delta,J}_B}{S^{\tilde\Delta,J}_{\Delta_0,\Delta_0}}\right)^{\frac{1}{2}} \,,
}
such that $f^\text{local}_{\Delta,J} f^\text{local}_{\tilde\Delta,J} = (-1)^J$, the quadratic term in the bulk action can be written in the simple local way:
\es{quadLoc}{
\mathcal{S}^{(2)}_\text{free}[\Phi_J] = \sum_{J=0}^\infty &
	 \frac{1}{\alpha_J}
	\int\frac{d^dxdz}{z^{d+1}} \Phi_{J}(x,z) \left(\nabla_{x,z}^{2}-M_{d+J-2,J}^{2}\right)\left(\nabla_{x,z}^{2}-M_{d+J,J}^{2}\right) \Phi_{J}(x,z)\,,
}
where for $d<4$ the modified map \eqref{J0} gives a slightly different form for the $J=0$ term as shown in \cite{Aharony:2020omh}. The higher order terms $\mathcal{S}^{(n)}[\Phi_J]$ in the bulk action remain non-local even for $f_{\Delta,J}^\text{local} $, and include explicit bulk counterterms starting with $\mathcal{S}^{(1)}[\Phi_J]$ that are dual to the $V$-dependent counterterms in \eqref{JacEta}.

The various deformations to the free theory discussed in Section \ref{sec:bilocal_QED} can then be mapped to the bulk using the off-shell map of local operators in \eqref{toshow}. For instance, the scalar double-trace deformation in \eqref{actionEta2} maps to
\es{lamBulk}{
\mathcal{S}_\lambda[\Phi_0,\sigma]=\int {d^dx}\left[-\frac{1}{4\lambda}\sigma^2(x)+\frac{1}{2}\sigma(x)\frac{1}{f_{d-2,0}}\lim_{\eps\to0}\eps^{2-d}\Phi_0(x,\eps)\right]\,,
}
where $\sigma(x)$ can be thought of as living on the boundary of AdS. When $\lambda\to0$, this defines the bulk dual of the critical $U(N)$ theory for any $d>2$ such that this CFT exists, which was argued to be (for large enough $N$) $2<d<6$ in \cite{Fei:2014yja}\footnote{The CFT is believed to be unitary only for $2<d<4$.}. The $\sigma(x)$ field then acts as a Lagrange multiplier in \eqref{lamBulk} that sets $\lim_{\eps\to0}\eps^{2-d}\Phi_0(x,\eps)=0$ off-shell, just as in the CFT it set $\eta(x,x)=0$ off-shell. Thus, the off-shell relation \eqref{toshow2} becomes trivial in the critical theory. The vanishing of the $z^{d-2}$ mode implies that $\Phi_0(x,z)$ now has the same small $z$ boundary condition as all other $J>0$ bulk fields, namely it scales as $z^{d/2}$, which is the real part of the principal series contour. 
For $2<d<4$ this was already discussed in \cite{Aharony:2020omh}, and here we can generalize this to $d>4$ because we generalized the off-shell map for $J=0$ to $d>4$.

For QED, we first gauge-fix the CFT by adding the term $\mathcal{S}_\zeta[A]$ given in \eqref{GFaction}, where the family of possible gauge-fixings is parameterized by $\zeta\in\mathbb{R}$. Since this term, as well as the Chern-Simons term $\mathcal{S}_k[A]$ given in \eqref{CSaction} for $d=3$, do not depend on $\eta$, they map trivially to the bulk such that $A_i(x)$ now lives on the boundary of AdS, just like $\sigma$. We then use the off-shell map \eqref{toshow} for $J=0,1$ to map $\mathcal{S}_e[\eta,A]$ to the bulk to get
\es{eBulk}{
\mathcal{S}_e[\Phi_0,\Phi_1,A]&=\int {d^dx}\Bigg[\frac{1}{4e^2}F_{ij}^2(x)+A_i^2(x)G_0(x,x)\\
&\qquad\qquad\quad+\lim_{\eps\to0}\eps^{2-d}\Big[\frac{2i\sqrt{d-2}}{f_{d-1,1}}A_i(x)\Phi_1^i(x,\eps)+\frac{A_i^2(x)\Phi_0(x,\eps)}{\sqrt{N}f_{d-2,0}}\Big]\Bigg]\,,
}
where we identified $S^1_i(x)=-\frac{i}{2\sqrt{d-2}}J_i[\eta]$, and recall that the counterterm $A_i^2(x)G_0(x,x)$ is necessary to cancel divergences in the $A_i$ two-point function. We can then take the limit $e\to\infty$ to get the bulk dual scalar QED \eqref{CPNS} with bulk action:
\es{bulkQEDs}{
\mathcal{S}_{\mathbb{CP}^{N-1}}[\Phi_J,A,\sigma]&\equiv\sum_{n=1}^\infty\mathcal{S}_\text{free}^{(n)}[\Phi_J]+ \mathcal{S}_{e\to\infty}[\Phi_0,\Phi_1,A]+\mathcal{S}_{\lambda\to\infty}
}
as well as the Chern-Simons term $\mathcal{S}_k[A]$ for $d=3$ (if desired).

Similar to the $\sigma, \Phi_0$ case, we can think of $A_i(x)$ as a Lagrange multiplier that for $k=0$ sets $\lim_{\varepsilon\to0}\varepsilon^{2-d}\Phi_1(x,\varepsilon)=0$ as an operator equation, just as in the CFT it set to zero the $U(1)$ current. The $z^0$ mode of $\Phi_1$ then becomes dynamical, according to the off-shell relation:
\es{offshellspin1}{
\text{QED with $k=0$}:\qquad \Phi_1^i(x,0) = i 2\sqrt{d-2} \mathfrak{a}_1 f_{d-1,1} S_B^{d-1,1}A_i(x)\,,
}
which is just \eqref{bulkBound} with $J=1$ and $S^1_i(x)=-\frac{i}{2\sqrt{d-2}}J_i[\eta]$. Namely, in the bulk dual to QED, $A_i(x)$ is nothing but the boundary value of the bulk vector field. It is convenient to write this change of boundary condition by defining a boundary field strength for $\Phi_1$ (even though in our formalism there is no gauge freedom for this field) as 
\es{F_def}{
  \mathcal{F}_{\mu\nu}(x) \equiv \lim_{z\to0}\left(\partial_\mu \Phi_{1,\nu}(x,z)-\partial_\nu \Phi_{1,\mu}(x,z)\right)\,.
}
The boundary conditions for the free scalar theory and for QED for general $d$, as well as for $d=3$ and general $k$, can then be written compactly as
\es{boundF}{
&\text{free}:\qquad \qquad\qquad\qquad \mathcal{F}_{ij}(x)=0\,,\\
&\text{QED$_d$ with $k=0$}:\qquad \mathcal{F}_{zi}(x)=0\,,\\
&\text{QED$_3$}:\qquad\qquad\qquad\quad \mathcal{F}_{ij}(x)+\frac{2\pi i}{16\kappa} \varepsilon_{ijl} \mathcal{F}_{zl}(x) =0\,,\\
}
where the free theory corresponds to electric boundary conditions, QED with $k=0$ corresponds to magnetic boundary conditions, and QED$_3$ for general $k$ corresponds to mixed boundary conditions. Finally, the $A_i^2(x)$ term in \eqref{eBulk} will also alter the boundary behavior of $\Phi_0$ at subleading order in $1/N$, according to \eqref{toshow2}.

\subsection{The bulk correlation functions}
\label{sec:correlations}

We will now discuss the bulk correlation functions that follow from the bulk actions written above. For the dual of the free theory with $\lambda=e=0$, the bulk two-point functions coming from the local quadratic action \eqref{quadLoc} are \cite{Aharony:2020omh}:
\es{freeBulk2}{
\langle & \Phi_J (x_1,z_1) \Phi_J(x_2,z_2)\rangle = \frac{\alpha_J/2}{M^2_{d+J,J}-M^2_{d+J-2,J}} 
			\left(
				\Pi^{TT}_{d-2+J,J}(x_1,z_1|x_2,z_2)-\Pi^{TT}_{d+J,J}(x_1,z_1|x_2,z_2)
			\right)\,.
}
Here, $\Pi^{TT}_{d-2+J,J}(x_1,z_1|x_2,z_2)$ are the traceless transverse bulk-to-bulk propagators defined in \cite{Aharony:2020omh} by the differential equation
\es{eq:bulk_bulk_DE_2}{
 	\left( \nabla^2_{x,z} - M^2_{\Delta,J}\right) \Pi^{TT}_{\Delta,J}(x_1,z_1|x_2,z_2){}   = -\delta^{TT}(x_1,z_1|x_2,z_2)\,,
}
and by the $z_2\to0$ boundary condition
\es{eq:BB_limit}{
	\Pi^{TT}_{\Delta,J}(x_1,z_1|x_2,z_2) =
	\begin{cases}
{z_2}^{\Delta-J} {\cal C}_{\Delta,J} G_{\Delta,J}(x_1,z_1|x_2) +O(z_2^{\Delta-J+1}) \qquad \Delta<d-J+4\\
O(z_2^{d-2J+4})\qquad\qquad\qquad\qquad\qquad\qquad\quad\;\;\;\, \Delta\geq d-J+4\,, \\
\end{cases}
}
where $\delta^{TT}$ is a delta function for traceless transverse functions that is defined precisely in \cite{Aharony:2020omh}, and the normalization
is 
\es{eq:bb_limit_coeff}{
	{\cal C}_{\Delta,J} \equiv \frac{(J+\Delta-1)\Gamma(\Delta-1)}{2\pi^{\frac{d}{2}} \Gamma(\Delta+1-\frac{d}{2})}\,.
}
The solution to \eqref{eq:bulk_bulk_DE_2} and \eqref{eq:BB_limit} is by construction traceless and transverse:
	\es{trans}{
\text{for $a=1,\cdots,J$:}\qquad&\nabla^{\mu_a} \Pi^{TT}_{\Delta,J}(x_1,z_1|x_2,z_2){}_{\mu_1,\dots, \mu_a,\dots,\mu_J|\mu'_1,...,\mu'_J}=0\,,\\
&  \Pi^{TT}_{\Delta,J}(x_1,z_1|x_2,z_2){}_{\mu_1,\dots, \mu_a,\dots,\mu_J|\mu'_1,\dots,\mu_a,\dots,\mu'_J}=0\,.
}
The explicit propagators for $\Delta\geq d-2+J$ for $J>0$, or $J=0$ and any $\Delta$, were computed in \cite{Aharony:2020omh}, and are reviewed in Appendix \ref{erez1}. The massless propagators in the first term of \eqref{freeBulk2} are identified with the physical particles dual to the operators in our theory\footnote{For massless propagators, the transversality constraint \eqref{trans} can be interpreted as a bulk gauge choice in a putative bulk gauge theory, which would be related to our bulk theory upon gauge-fixing. }. For $J>0$ they are dual to the infinite conserved currents in the CFT, and for $J=0$ to the $\Delta=d-2$ single-trace scalar. The particle with negative propagator and $\Delta=d+J-2$ is sub-leading in the $z\to0$ limit (it happens to match the ghost spectrum of a certain gauge-fixing of Vasiliev's theory \cite{Gaberdiel:2010ar,Gaberdiel:2010xv,Gupta:2012he}, so we expect that probably it can be interpreted as some kind of gauge-fixing-related ghost). The higher order correlation functions can be computed using the Feynman rules for the higher order terms $\mathcal{S}^{(n)}[\eta]$, which were shown in \cite{Aharony:2020omh} to lead to the expected free theory results, where all loops and counterterms cancel (this cancellation will no longer be exact in the non-free theory we are discussing here, but the divergences should cancel in the same way).

We can then consider the effect of the quadratic scalar deformation $\mathcal{S}_{\lambda}[\Phi_0,\sigma]$ given in \eqref{lamBulk}, which only modifies the scalar propagator as \cite{Aharony:2020omh}:
\es{bulkCrit2p}{
&\langle  \Phi_0(x_1,z_1)\Phi_0(x_2,z_2\rangle_{\lambda}=\langle  \Phi_0(x_1,z_1)\Phi_0(x_2,z_2)\rangle_{\lambda=0}+\frac{1}{4(f_{d-2}^\text{local})^2}\left[\frac{\alpha_0\mathcal{C}_{d-2,0}/2}{M^2_{d,0}-M^2_{d-2,0}} \right]^2\\
&\qquad\qquad \times \int d^dyd^dy' G_{d-2,0}(x_1,z_1|y)\langle\sigma(y)\sigma(y')\rangle_\lambda G_{d-2,0}(x_2,z_2|y')+O(N^{-1})\,,
}
where the $\sigma$ propagator at finite $\lambda$ is given in \eqref{sigProp}. For the critical theory at $\lambda\to\infty$, the $\sigma$ propagator in \eqref{sigProp2} becomes a conformally invariant two-point function with scaling dimension $\Delta=2+O(1/N)$, so we can use the identity \cite{Hartman:2006dy,Giombi:2011ya,Costa:2014kfa} 
\es{oldIdentity}{
&\int d^dyd^dy' {G_{\Delta,0}(x_1,z_1|y) \langle\mathcal{O}_{\tilde\Delta,0}(y)\mathcal{O}_{\tilde\Delta,0}(y')\rangle G_{\Delta,0}(x_2,z_2|y')}\\
&\qquad=\alpha_0 N_{\Delta,0}\left(\Delta-\frac d2\right)\left(\Pi^{TT}_{\Delta,0}(x_1,z_1;x_2,z_2)-\Pi^{TT}_{\tilde\Delta,0}(x_1,z_1;x_2,z_2)\right)\,,
}
with $\Delta=d-2$ to compute
\es{bulkCrit2p2}{
&\langle  \Phi_0(x_1,z_1)\Phi_0(x_2,z_2)\rangle_{\lambda\to\infty}=\frac{\alpha_0/2}{M_{d,0}^2-M_{d-2,0}^2}(\Pi^{TT}_{2,0}(x_1,z_1;x_2,z_2)-\Pi^{TT}_{d,0}(x_1,z_1;x_2,z_2))+O(N^{-1})\,.\\
} 
We see that the physical propagator $\Pi^{TT}_{d-2,0}$ for $\Phi_0$ in the bulk dual of the free scalar theory has been replaced by the shadow propagator $\Pi^{TT}_{2,0}$, which generalizes the $2<d<4$ results of \cite{Aharony:2020omh} to $2<d<6$ (where the critical theory is defined at large $N$). Both $\Pi^{TT}_{\Delta,J}$ and $\Pi^{TT}_{\tilde\Delta,J}$ are defined by the same bulk differential equation \eqref{eq:bulk_bulk_DE_2}, and differ only by the boundary condition \eqref{eq:BB_limit}. Since the only difference between the free and critical bulk theories was this scalar propagator, we see that to all orders in $1/N$ the only difference between the free and critical bulk theories is the boundary condition for the bulk scalar, as anticipated in \cite{Klebanov:1999tb,Giombi:2011ya,Hartman:2006dy}. As shown in \cite{Giombi:2011ya}, the modification of the free bulk Feynman rules by replacing $\Pi^{TT}_{d-2,0}\to \Pi^{TT}_{2,0}$ then leads to the expected bulk dual of the critical CFT for all bulk correlators, at all orders in $1/N$.

Next, we consider the effect of coupling to the gauge field by adding $\mathcal{S}_{e}[\Phi_0,\Phi_1,A]$ with the family of gauge-fixing terms $\mathcal{S}_{\zeta}[A]$ (and for $d=3$ we can add also the Chern-Simons term $\mathcal{S}_{k}[A]$). The $J=1$ propagator is modified as
\es{bulkJ1Mod}{
&\langle  \Phi_1(x_1,z_1)\Phi_1(x_2,z_2)\rangle_{\zeta,e,k}=\langle  \Phi_1(x_1,z_1)\Phi_1(x_2,z_2)\rangle_{e=0}+\frac{4(2-d)}{(f_{d-1,1}^\text{local})^2}\left[\frac{\alpha_1\mathcal{C}_{d-1,1}/2}{M^2_{d+1,1}-M^2_{d-1,1}} \right]^2\\
&\qquad\qquad \times \int d^dyd^dy' G^i_{d-1,1}(x_1,z_1|y)\langle A_i(y)A_j(y')\rangle_{\zeta,e,k} G^j_{d-1,1}(x_2,z_2|y')+O(N^{-1})\,,
}
where we suppressed the bulk indices for simplicity, and where the gauge-fixed photon propagator is given in \eqref{Aprop} for the $\zeta$ family of gauges.
Note that the bulk propagator now depends on the choice of gauge-fixing in the CFT. For the critical theory at $e\to\infty$, we can derive a generalization of the identity \eqref{oldIdentity} to massless $J=1$ propagators:\footnote{A similar identity in axial gauge was derived in \cite{Chang:2012kt}.}
\es{identityNew}{
&\int d^dyd^dy' {G_{d-1,1}(x_1,z_1|y){}_{i|\mu} \langle A_i(y)A_{i'}(y') \rangle_{\zeta,e\to\infty,k} G_{d-1,1}(x_2,z_2|y')}{}_{i'|\mu'}\\
&\qquad=\frac{8\pi^\frac{3d}{2}}{\Gamma\left(\frac{d-2}{2}\right)\Gamma\left(d\right)}
\left(\Pi^{TT}_{d-1,1}(x_1,z_1;x_2,z_2)_{\mu|\mu'}-\Pi^{\zeta,k}_{1,1}(x_1,z_1;x_2,z_2)_{\mu|\mu'}\right)\,.
}
Here, $\Pi^{\zeta,k}_{1,1}$ is a traceless and transverse propagator that satisfies the same differential equation \eqref{freeBulk2} as $\Pi^{TT}_{d-1,1}$, but has the alternate boundary condition
\es{newBound}{
\lim_{z_2\to0}\Pi^{\zeta,k}_{1,1}(x_1,z_1;x_2,z_2)= \frac{\Gamma\left(\frac{d-2}{2}\right)\Gamma^{2}\left(\frac{d}{2}\right)}{2\pi^{\frac{d}{2}}\Gamma\left(d\right)} G^{\zeta,k}_{1,1}(x_1,z_1|x_2)\,,
}
where $G^{\zeta,k}_{1,1}(x_1,z|x_2)$ is a transverse bulk-to-boundary propagator that satisfies the same differential equation \eqref{eq:bulk_boundary_DE} as $G_{d-1,1}(x_1,z|x_2)$ but has the alternate $z\to0$ boundary condition
\es{newBounds2}{
G^{\zeta,k}_{1,1}(x_1,z|x_2)_{ij}=\begin{cases} \langle A_i(x_1)A_j(x_2)\rangle_{\zeta,\infty}-z^{d-2}\frac{\pi^{\frac{d}{2}}\Gamma\left(d\right)}{\Gamma^{3}\left(\frac{d}{2}\right)}\int\frac{d^dp}{(2\pi)^d}e^{ipx_{12}}(\delta_{ij}-\frac{p_ip_j}{|p|^2})+\dots\,,\qquad d\neq3\,,\\
 \langle A_i(x_1)A_j(x_2)\rangle_{\zeta,\infty,k}-z\int\frac{d^{3}p}{\left(2\pi\right)^{3}}e^{ip x_{12}}\frac{16\left(\delta_{ij}-\frac{p_{i}p_{j}}{\left|p\right|^{2}}\right)-\frac{128\kappa}{\pi}\,\varepsilon_{ijk}\frac{p_{k}}{\left|p\right|}}{\frac{64\kappa^2}{\pi^2}+1}+\dots\,,\quad d=3\,.\\
\end{cases}
}
Note that we expressed $\delta^{TT}_1(x_{12})$ explicitly in momentum space for $d\neq3$ for ease of comparison to $d=3$. Alternatively, $\Pi_{1,1}^{\zeta,k}$ can be defined by their differential equation and the boundary condition \eqref{boundF}.
Explicit expressions for $\Pi^{\zeta,k}_{1,1}$ and $G^{\zeta,k}_{1,1}$ are derived in Appendix \ref{erez2}, where we show that they satisfy \eqref{identityNew}. We can then apply \eqref{identityNew} to the $e\to\infty$ limit of \eqref{bulkJ1Mod} to compute
\es{bulkCrit2p2A}{
&\langle  \Phi_1(x_1,z_1)\Phi_1(x_2,z_2)\rangle_{\zeta,e\to\infty,k}=\frac{\alpha_1/2}{M_{d+1,1}^2-M_{d-1,1}^2}(\Pi^{\zeta,k}_{1,1}(x_1,z_1;x_2,z_2)-\Pi^{TT}_{d+1,1}(x_1,z_1;x_2,z_2))+O(N^{-1})\,.\\
} 
As in the scalar deformation, we see that the physical propagator $\Pi^{TT}_{d-1,1}$ for $\Phi_1$ has been replaced by the shadow propagator $\Pi^{\zeta,k}_{1,1}$, except that now the shadow propagator depends on the CFT gauge-fixing parameter $\zeta$, and for $d=3$ it can also have an infinite number of possible boundary conditions, parameterized by $k$. 

For the $\mathbb{CP}^{N-1}$ model, we saw above that we can change variables for $\sigma(x)$ to get rid of the $A_i^2 \phi^2$ term. In this way $\sigma,A_i$ couple in the bulk action \eqref{bulkQEDs} only linearly to $\Phi_0,\Phi_1$ respectively. As we explained in this section, this means that the $\mathbb{CP}^{N-1}$ bulk dual (at any order in $1/N$) is the same as the free theory bulk dual, only with the alternative boundary conditions for both $\Phi_0$ and $\Phi_1$. We can use this to argue that our bulk action gives the expected correlators to all orders in $1/N$ for the $\mathbb{CP}^{N-1}$ model, following the analogous argument for the critical $O(N)$ model in \cite{Giombi:2011ya}. In particular, already in the local description \eqref{CPNS} we could change variables for $\sigma$, which led to only linear couplings for $\sigma,A_i$ to the $\phi_I$'s. We can then consider the difference between correlation functions of single-trace operators in the $\mathbb{CP}^{N-1}$ model and in the free theory. At any order in $1/N$, this difference can be written as integrals of the free theory $S^0, S^1_i$ correlation functions with the effective propagators for $\sigma,A_i$. Using \eqref{oldIdentity} for $S^0$ and \eqref{identityNew} for $S^1_i$, we can write the difference between the correlators as the difference between the same Witten diagrams, only with alternative boundary conditions for $\Phi_0$ and $\Phi_1$. 


\section{The bulk dual of $U(N_c)$ scalar QCD at finite $N_c$} 
\label{sec:QCD_bulk}

The bulk dual of scalar QED in $2<d<4$ with $N\gg1$ scalars that we described in the previous sections can be easily generalized to $U(N_c)$ QCD in $2<d<4$ with $N\gg1$ scalars \cite{PhysRevLett.64.721}, for finite $N_c$. For $d=3$ we can again consider also the large Chern-Simons level $k$ limit with fixed $\kappa\equiv k/N$. 

We start by considering a free theory of $N\times N_c$ scalars with global symmetry $U(NN_c)$. The singlet sector under the $U(N)\subset U(NN_c)$ subgroup can be translated to the bulk just like the usual $U(N)$ free scalar theory, except that now both the $U(N)$ singlet bi-locals in the CFT and their dual bulk fields are adjoints under $U(N_c)$, and the terms in the bulk action are all single traces of products of $U(N_c)$ matrices. We can then construct the bulk dual of scalar QCD by a procedure similar to the one described in the previous sections: gauging $U(N_c)$ (in the CFT), fixing a gauge convenient for the large $N$ expansion, mapping the $U(N_c)$ current (which is a $U(N)$ singlet) to the bulk using our usual CFT-to-AdS map, and coupling it to the new $U(N_c)$ gauge fields. The main differences from the Abelian case are that now the Yang-Mills and Chern-Simons terms for $U(N_c)$ contain self-interactions of the $U(N_c)$ gauge fields, and that the gauge-fixing now leads to ghosts (in the adjoint of $U(N_c)$) which will live on the boundary and couple to the $U(N_c)$ gauge fields there.

As with QED, the resulting bulk theory is related to the bulk dual of the free theory by a simple change of boundary conditions for the spin one bulk field (in the adjoint of $U(N_c)$). Note that this bulk construction is useful for large $N$ with fixed $N_c$, and it does not apply to the limit of large $N_c,k$ with fixed $N,k/N_c$ discussed in \cite{Aharony:2011jz,Giombi:2011kc}, which is expected to have a different dual description.

\section*{Acknowledgments}

We would like to thank Tal Sheaffer and Tomer Solberg for collaboration on related issues.
This work was supported in part  by an Israel Science Foundation center for excellence grant (grant number 2289/18), by grant no. 2018068 from the United States-Israel Binational Science Foundation (BSF), and by the Minerva foundation with funding from the Federal German Ministry for Education and Research. OA is the Samuel Sebba Professorial Chair of Pure and Applied Physics. SMC is supported in part by a Zuckerman STEM leadership fellowship.

\pagebreak
\appendix

\section{Conformal bases} \label{bases}

In this appendix we review properties of the conformal bases that we use for the CFT and for the bulk fields. For the CFT we follow \cite{Dobrev:1976vr,Dobrev:1977qv}, while for the bulk we follow \cite{Costa:2014kfa}. 

\subsection{Three-point function basis}
\label{3pointBasis}

The harmonic basis defined in \eqref{3point} satisfies the completeness relation
\es{complete}{
\delta(x_{13})\delta(x_{24})=\frac12\sum_{J=0}^\infty \int_{\gamma_J}\frac{d\Delta}{2\pi i} \int d^d y  \, 
	\frac{1}{N_{\Delta,J}}
	\thpnorm{x_1}{x_2}{\Delta,J}{y}\thptild{x_3}{x_4}{\tilde\Delta,J}{y}\,,
}
where the contour $\gamma_J$ was described in the main text, and the normalization factor is
\es{CFTNorm}{
	N_{\Delta,J}=\frac{\pi^{\frac{3d}{2}}\Gamma(J+1)}
	{2^{J-1}\Gamma(\frac{d}{2}+J)}
	\frac{\Gamma\left(\Delta-\frac{d}{2}\right)}{\Gamma\left(\Delta-1\right)\left(\Delta+J-1\right)}\frac{\Gamma\left(\tilde{\Delta}-\frac{d}{2}\right)}{\Gamma\left(\tilde{\Delta}-1\right)\left(\tilde{\Delta}+J-1\right)}\,.
	}
The basis also satisfies the orthogonality relation
\es{ortho}{
	&\int d^d x_{1}d^d x_{2}
	\thpnorm{x_1}{x_2}{\Delta,J}{y}\thptild{x_1}{x_2}{\tilde\Delta^\prime,J^\prime}{y^\prime}\\
	& \quad = 2\pi i \,\, N_{\Delta,J}\left( \delta_{J,J^\prime}\delta\left(\Delta-\Delta^\prime\right)\delta^{TT}_J(y-y^\prime) + \frac{\delta\left(\Delta-\tilde \Delta^\prime\right)}{S^{(\tilde\Delta,J)}_{\Delta_0,\Delta_0} } \left\langle O_{\Delta,J}(y) O_{\Delta,J'}(y^\prime) \right\rangle\right)\,,
	}
where the traceless transverse delta function $\delta^{TT}_J$ was defined in the main text, and the shadow coefficient is
\es{S}{
	S^{(\tilde\Delta,J)}_{\Delta_0,\Delta_0} = \frac{\pi^{\frac{d}{2}} \Gamma\left( \tilde \Delta -\frac{d}{2} \right)\Gamma\left(\tilde\Delta+J-1\right)\Gamma^2\left(\frac{\Delta+J}{2}\right)}
	{\Gamma\left(\tilde\Delta-1\right)\Gamma\left(\Delta+J\right)\Gamma^2\left(\frac{\tilde\Delta+J}{2}\right)}\,.
}
The reason for the second term in \eqref{ortho} is because the basis elements for $\Delta$ and for $\tilde \Delta$ are related by the shadow transform:
\begin{equation}
	\thpnorm{x_1}{x_2}{\Delta,J}{y}
	= \frac{1}{S^{(\tilde\Delta,J)}_{\Delta_0,\Delta_0} } 
	\int d^d y^\prime 
	\tp{\Delta,J}{y}{y^\prime} \thpnorm{x_1}{x_2}{\tilde\Delta,J}{y^\prime}\label{eq:shadow_trans}\,.
\end{equation}

\subsection{Bulk-to-boundary propagator basis}
\label{bulkBasis}

The differential equation \eqref{eq:bulk_boundary_DE} with boundary condition \eqref{eq:2_point_limit} has the explicit solution
\es{explicit_prop}{
	G _{\Delta,J}(x,z|y){}_{\mu_1,...,\mu_J|i_1,..,i_J}= \left(\frac{z}{(x-y)^2+z^2}\right)^\Delta \left(X_{\mu_1,i_1}\cdot...\cdot X_{\mu_J,i_J}-\text{traces}\right)\,,
}
where $X_{i,j} = z^{-1}\Big({\delta_{i,j}-2\frac{(x-y)_i (x-y)_j}{(x-y)^2+z^2}}\Big)$, $X_{z,i}=\frac{-2(x-y)_i}{(x-y)^2+z^2}$. The propagator satisfies the bulk analog of the shadow transform \eqref{eq:shadow_trans}: 
\es{eq:shadow_trans_bulk}{
	& G^{i_1 \dots i_J}_{\Delta,J}(x,z|y  )_{\mu_1,...,\mu_J}  =  \frac{1}{S_B^{\tilde \Delta,J}}\int d^dy' G^{i'_1 \dots i'_J}_{\tilde\Delta,J}(x,z|y'  )_{\mu_1,...,\mu_J} 
	\langle\mathcal{O}_{\Delta,J}^{i_1\dots i_J}(y')\mathcal{O}_{\Delta,J}^{i'_1\dots i_J'}(y')\rangle \,,
}
where $S_B^{\tilde \Delta,J}$ was given in \eqref{eq:bulk_shadow_coeff}. The orthogonality relation takes the form
\es{Bortho}{
&\int\frac{d^dxdz}{z^{d+1}} G^{\mu_1\dots\mu_J|i_1\dots i_J}_{\Delta,J}(x,z|x_1)G^{i'_1\dots i'_J}_{\tilde\Delta',J}(x,z|x_2)_{\mu_1\dots\mu_J}=\\
& 2\pi i\delta(\Delta-\tilde\Delta')\langle\mathcal{O}^{i_1\dots i_J}_{\Delta,J}(x_1)\mathcal{O}^{i'_1\dots i'_J}_{\Delta,J}(x_2)\rangle S_B^{\tilde\Delta,J}+2\pi i \delta(\Delta-\Delta')\delta^{TT}_{J,{i_1\dots i_j|i'_1\dots i'_J}}(x_{12})\frac{N_{\Delta,J}\alpha_J}{2}\,,
}
where $\delta^{TT}_J(x)$ is defined in \cite{Costa:2014kfa} as a delta function for traceless transverse spin $J$ functions in $d$ dimensions, $N_{\Delta,J}$ is defined in \eqref{CFTNorm}, and
\es{eq:alpha_J_def}{
 	\alpha_J  \equiv \frac{2^J \Gamma(\frac{d}{2}+J)}{\pi^{\frac{d}{2}}\Gamma(J+1)}\,.
 	} 
The bulk completeness relation for traceless transverse bulk functions is 
\begin{equation}
\delta^{\mu_1,...,\mu_J|\mu'_1,...,\mu'_J}_{TT}(x,z|x',z')=\int_{P.S.}\frac{d\Delta}{2\pi i} \int d^dy \frac{G^{\mu_1,...,\mu_J|i_1,..,i_J}_{\Delta,J}(x,z|y)G^{\mu'_1,...,\mu'_J|i_1,..,i_J}_{\tilde{\Delta},J}(x',z'| y  )}{\alpha_J N_{\Delta,J}}
	\,,\label{eq:omega_def}
\end{equation}
where here $\delta^{TT}$ is defined in the bulk.

\section{Bulk-to-bulk propagators}
\label{erez}

In this appendix, we will show explicit expressions for the bulk-to-bulk propagators discussed in the main text. We will start by reviewing the traceless transverse propagators $\Pi^{TT}_{\Delta,J}$ introduced in \cite{Aharony:2020omh}, which are defined by the differential equation \eqref{eq:bulk_bulk_DE_2} with standard boundary condition \eqref{eq:BB_limit}. These have explicit position space expressions for all $J$ and $\Delta\geq d-2+J$. For $J=1$, we will also give the explicit expression in momentum space. We will then consider the $J=1$ propagator defined by the same differential equation \eqref{eq:bulk_bulk_DE_2} but with the alternate boundary condition \eqref{newBound}, which we will also write explicitly in momentum space. From these momentum space expressions, we can immediately see the identity \eqref{identityNew}.

\subsection{Standard boundary conditions}
\label{erez1}

The differential equation \eqref{eq:bulk_bulk_DE_2} with boundary condition \eqref{eq:BB_limit} can be formally solved using the split representation
\es{usprop}{
&	\Pi^{TT}_{\Delta,J}(x,z|x',z')=  \int_{\gamma_J} \frac{d \Delta^\prime}{2\pi i}\frac{G_{\Delta,J}(x,z|y)G_{\tilde{\Delta},J}(x',z'| y  )}{\alpha_JN_{\Delta,J}(M^2_{\Delta,J}-M^2_{\Delta^\prime,J})}\,,
	}
	where the boundary spin indices are contracted on the right-hand side, while the bulk indices on both sides are implicit. For massive propagators, i.e. $\Delta>d-2+J$ or $J=0$, we can close the contour and collect poles to get the explicit position-space expression
	\es{usprop2}{
	\Pi^{TT}_{\Delta,J}(x,z|x',z')  =\Pi_{\Delta,J}(x,z|x',z')-\sum_{p=d-1}^{d+J-2}  \frac{(2p-d)}{M^2_{\Delta,J}-M^2_{p,J}}\text{Res}\left[  \Pi_{\Delta^\prime,J}(x,z|x',z')   \right]_{\Delta^\prime=p}\,,
	}
where $\Pi_{\Delta,J}$ is the standard massive bulk-to-bulk propagator whose explicit form is given by a complicated recursion relation in \cite{Penedones:2010ue}. For instance, the $J=0$ propagator is
\es{J0prop}{
\Pi_{\Delta,0}(x,z|x',z')=\mathcal{C}_{\Delta,0}(2u)^{-\Delta}{}_2F_1\left(\Delta,\Delta+\frac{1-d}{2},2\Delta-d+1;-\frac{2}{u}\right)\,,
}
where the chordal distance is 
\es{u}{
u=\frac{\sum_ix_ix'_i +zz'}{2z z'}\,,
}
and note that $\Pi^{TT}_{\Delta,0}=\Pi_{\Delta,0}$ since transversality and tracelessness are trivial for $J=0$. For $J=1$ the propagator is 
\es{J1prop}{
\Pi^{TT}_{\Delta,1}(x,z|x',z'){}_{\mu|\mu'} =-g_0(u){\partial_\mu\partial_{\mu'}}u +g_1(u)\partial_{\mu}u\partial_{\mu'}u \,,
}
where $g_0(u)$ and $g_1(u)$ are
\es{J1prop2}{
g_0(u)&=\frac{ (d-\Delta ) \Gamma (\Delta +1) u^{-\Delta } \, _2F_1\left(\Delta ,\frac{1}{2} (-d+2 \Delta +1);-d+2 \Delta
   +1;-\frac{2}{u}\right)}{\pi ^{d/2} 2^{\Delta +1}(\Delta -1) (d-\Delta -1) \Gamma \left(-\frac{d}{2}+\Delta +1\right)}\\
   &-\frac{(u+1) \Gamma (\Delta +1)
   u^{-\Delta -1} \, _2F_1\left(\Delta +1,\frac{1}{2} (-d+2 \Delta +1);-d+2 \Delta +1;-\frac{2}{u}\right)}{\pi ^{d/2} 2^{\Delta +1}(\Delta -1) (d-\Delta -1) \Gamma
   \left(-\frac{d}{2}+\Delta +1\right)}\,,\\
g_1(u)&=\frac{ (u+1) (d-\Delta ) \Gamma (\Delta +1) u^{-\Delta -1} \, _2F_1\left(\Delta ,\frac{1}{2} (-d+2 \Delta +1);-d+2 \Delta
   +1;-\frac{2}{u}\right)}{\pi ^{d/2} 2^{\Delta +1}(\Delta -1) (u+2) (d-\Delta -1) \Gamma \left(-\frac{d}{2}+\Delta +1\right)}\\
   &-\frac{
   \left(d+(u+1)^2\right) \Gamma (\Delta +1) u^{-\Delta -2} \, _2F_1\left(\Delta +1,\frac{1}{2} (-d+2 \Delta +1);-d+2 \Delta +1;-\frac{2}{u}\right)}{\pi ^{d/2} 2^{\Delta +1}(\Delta
   -1) (u+2) (d-\Delta -1) \Gamma \left(-\frac{d}{2}+\Delta +1\right)}\,.\\
}
	In the massless limit $\Delta\to d-2+J$, the $\Delta'=d-2+J$ pole in \eqref{usprop} becomes a double pole and we get the finite result
	 \es{usprop3}{
	&\Pi^{TT}_{d-2+J,J}(x,z|x',z') =\\
	&\qquad \partial_\Delta\Big[(\Delta-d-J+2)\Pi_{\Delta,J}(x,z|x',z')\Big]_{\Delta=d-2+J}- \frac{\text{Res}   \left[   \Pi_{\Delta^\prime,J}(x,z|x',z') \right]_{\Delta^\prime=d-(2-J)}}{4-d-2J} \\
	&\qquad -\sum_{p=d-1}^{d+J-3}  \frac{(2p-d)}{M^2_{\Delta,J}-M^2_{p,J}}\text{Res}\left[  \Pi_{\Delta^\prime,J}(x,z|x',z')   \right]_{\Delta^\prime=p}\,.
	}
	For $J=1$, the massless propagator $\Pi^{TT}_{d-1,1}$ is the same as the position space Landau gauge propagator given in \cite{DHoker:1998bqu}, which can be explicitly checked from the definitions given here. 
	
	We will find it convenient to express $\Pi^{TT}_{d-1,1}$ in momentum space. Instead of directly transforming the known position space expression, we can instead solve the differential equation \eqref{eq:bulk_bulk_DE_2} in momentum space. We start by writing the bulk-to-boundary differential equation \eqref{eq:bulk_boundary_DE} in momentum space as
	\es{btBmom}{
	\left(z^{2}\partial_{z}^{2}+\left(3-d\right)z\partial_{z}-p^{2}z^{2}\right)G_{d-1,1}\left(p,z\right)_{i,j}-2izp_{i}G_{d-1,1}\left(p,z\right)_{z,j}&=0,\\
  \left(z^{2}\partial_{z}^{2}+\left(3-d\right)z\partial_{z}-p^{2}z^{2}+1-d\right)G_{d-1,1}\left(p,z\right)_{z,j}+2iz\sum_{i=1}^{d}p_{i}G_{d-1,1}\left(p,z\right)_{i,j}&=0,
	}
	with the $z\to0$ boundary condition \eqref{eq:2_point_limit}. We can solve this to get 
	  \es{stdmomprop}{
  G_{d-1,1}\left(p,z\right)_{i,j} & =\frac{\pi^{\frac{d}{2}}2^{2-\frac{d}{2}}}{\Gamma\left(d\right)}\left|pz\right|^{\frac{d}{2}-1}\left(z\frac{p_{i}p_{j}}{\left|p\right|}K_{\frac{d-4}{2}}\left(\left|p\right|z\right)+\delta_{i,j}\left(d-2\right)\,K_{\frac{d-2}{2}}\left(\left|p\right|z\right)\right)\\
  G_{d-1,1}(p,z)_{z,j} & =\frac{\pi^{\frac{d}{2}}2^{2-\frac{d}{2}}}{\Gamma\left(d\right)}\left|pz\right|^{\frac{d}{2}-1}\cdot\left(ip_{j}z\right)K_{\frac{d-2}{2}}\left(|p| z\right) \,,
  }
  which is the momentum space version of \eqref{explicit_prop}. Next, we use transversality to rewrite \eqref{eq:bulk_bulk_DE_2} as
	\es{photoneq}{
  \partial_{\mu}\left(z^{3-d}\left(\partial_{\mu}\Pi_{\nu\rho}-\partial_{\nu}\Pi_{\mu\rho}\right)\right)& =-\delta_{\nu,\rho}\delta(x,z|x',z')+z^{-d-1}\partial_{\rho'}z^{2}\partial_{\nu}\Pi_{d,0},\\
  \Pi_{d,0}\left(x,z|x',z'\right)&=\left(zz'\right)^{\frac{d}{2}}\int\frac{d^{3}p}{\left(2\pi\right)^{3}}e^{ip\cdot\left(x-x'\right)}\begin{cases}
I_{\frac{d}{2}}\left(\left|p\right|z\right)K_{\frac{d}{2}}\left(\left|p\right|z'\right) & z<z'\\
I_{\frac{d}{2}}\left(\left|p\right|z'\right)K_{\frac{d}{2}}\left(\left|p\right|z\right) & z'<z
\end{cases}\,,
}
which we then write in momentum space as
\es{pleasewrite}{
 \partial_{z}\left(z^{-d+3}\left(\partial_{z}\Pi_{d-1,1}\left(p,z,z'\right)_{i,j}-ip_{i}\Pi_{d-1,1}\left(p,z,z'\right)_{z,j}\right)\right)&\\
 -z^{-d+3}\left(p^{2}\Pi_{d-1,1}\left(p,z,z'\right)_{i,j}+p_{i}\sum_{k=1}^{d}p_{k}\Pi_{d-1,1}\left(p,z,z'\right)_{k,j}\right)
 &\\
 =-\delta_{i,j}\delta\left(z-z'\right)-z^{-d+1}p_{i}p_{j'}&\Pi_{d,0}\left(p,z,z'\right),
}
\es{pid}{
   \Pi_{d,0}\left(p,z,z'\right)&=\left(zz'\right)^{\frac{d}{2}}\begin{cases}
  I_{\frac{d}{2}}\left(pz\right)K_{\frac{d}{2}}\left(pz'\right) & z<z'\\
  I_{\frac{d}{2}}\left(pz'\right)K_{\frac{d}{2}}\left(pz\right) & z'<z
  \end{cases}\,.
}
We can solve this equation along with the boundary condition \eqref{eq:BB_limit} and the explicit momentum space $G_{d-1,1}(p,z)$ to get
\es{stdbbprop}{
   \Pi_{d-1,1}^{TT}\left(p,z,z'\right)_{i,j}&=\left(zz'\right)^{\frac{d}{2}-1}K_{\frac{d-2}{2}}\left(pz'\right)I_{\frac{d-2}{2}}\left(pz\right)\left(\delta_{i,j}-\frac{p_{i}p_{j}}{p^{2}}\right)\\
   &+\frac{p_{i}p_{j}}{p^{4}}\left[\left(zz'\right)^{d-1}\partial_{z,z'}\left(\left(zz'\right)^{-d+1}\Pi^{TT}_{d-1,1}(p,z,z')_{z,z'}\right)\right],\\
   \Pi_{d-1,1}^{TT}\left(p,z,z'\right)_{i,z'}&=i\frac{p_{i}}{p^{2}}z^{d-1}\partial_{z}\left(z^{-d+1}\Pi^{TT}_{d-1,1}(p,z,z')_{z,z'}\right),\\
   \Pi_{d-1,1}^{TT}\left(p,z,z'\right)_{z,j}&=-i\frac{p_{j}}{p^{2}}\left(z'\right)^{d-1}\partial_{z'}\left(\left(z'\right)^{-d+1}\Pi^{TT}_{d-1,1}(p,z,z')_{z,z'}\right),\\
   \Pi_{d-1,1}^{TT}\left(p,z,z'\right)_{z,z'} &= -\frac{p^{2}}{d-2}\left(\partial_{\Delta}\Pi_{\Delta,0}\left(p,z,z'\right)\right)\mid_{\Delta=d-1}\,,
}
where $\Pi_{\Delta,0}\left(p,z,z'\right)$ is the scalar bulk-to-bulk propagator
\es{scalarbb}{
  \Pi_{\Delta,0}\left(p,z,z'\right)=\begin{cases}
\left(zz'\right)^{\frac{d}{2}}I_{\Delta-\frac{d}{2}}\left(\left|p\right|z\right)K_{\Delta-\frac{d}{2}}\left(\left|p\right|z'\right) & z<z'\\
\left(zz'\right)^{\frac{d}{2}}I_{\Delta-\frac{d}{2}}\left(\left|p\right|z'\right)K_{\Delta-\frac{d}{2}}\left(\left|p\right|z\right) & z>z'
\end{cases}.
}

\subsection{Alternate boundary conditions for $J=1$}
\label{erez2}

We will now solve the differential equation \eqref{pleasewrite} with the alternate boundary conditions \eqref{newBound}. We start by computing the alternate bulk-to-boundary propagator $G_{1,1}^{\zeta,k}(p,z)$, by solving \eqref{btBmom} with the boundary condition \eqref{newBounds2} and the explicit momentum space photon 2-point function given in the first line of \eqref{Aprop2}, which depends on the family of gauge-fixings parameterized by $\zeta\in\mathbb{R}$, as well as the Chern-Simons coupling $k$ for $d=3$. For $d\neq3$ we get 
  \es{altmomprop}{
    G^{\zeta}_{1,1}\left(p,z\right)_{i,j}&=\frac{2^{\frac{d}{2}-2}\pi^{\frac{d}{2}}\Gamma\left(d\right)\left|p\right|^{\frac{2-d}{2}}z^{\frac{d-2}{2}}}{\Gamma\left(2-\frac{d}{2}\right)\Gamma\left(\frac{d}{2}-1\right)\Gamma^{2}\left(\frac{d}{2}\right)}\left(\left(\delta_{i,j}-\frac{p_{i}p_{j}}{p^{2}}\right)K_{\frac{d-2}{2}}\left(\left|p\right|z\right)+\frac{4\left(1-\zeta\right)}{d-2}\frac{p_{i}p_{j}}{\left|p\right|}zK_{\frac{d}{2}}\left(\left|p\right|z\right)\right)\,,\\
    G^{\zeta}_{1,1}\left(p,z\right)_{z,j}&=i\frac{2^{\frac{d-2}{2}}\pi^{\frac{d}{2}}\Gamma\left(d\right)}{\Gamma\left(2-\frac{d}{2}\right)\Gamma^{3}\left(\frac{d}{2}\right)}\left(\zeta-1\right)\left|p\right|^{\frac{2-d}{2}}z^{\frac{d}{2}}K_{\frac{d-2}{2}}\left(\left|p\right|z\right)p_{j}\,,
  }
  while for $d=3$ and general $k$ we get
  \es{Bb3d}
{
  G_{1,1}^{\zeta,k}\left(p,z\right)_{i,j}&=\frac{e^{-\left|p\right|z}}{\left|p\right|}\frac{\frac{1}{16}\left(\delta_{i,j}+\frac{p_{i}p_{j}}{\left|p\right|^{2}}\left(\left(1-\zeta\right)\left(1+pz\right)-1\right)\right)-\frac{\kappa}{2\pi}\,\varepsilon_{ijk}\frac{p_{k}}{\left|p\right|}}{\left(\frac{\kappa}{2\pi}\right)^{2}+\left(\frac{1}{16}\right)^{2}}\\
  G_{1,1}^{\zeta,k}\left(p,z\right)_{z,j}&=i\frac{16\left(1-\zeta\right)}{1+\left(\frac{16\kappa}{2\pi}\right)^{2}}e^{-\left|p\right|z}\frac{z}{\left|p\right|}\,p_{j}\,.
}
We can then solve \eqref{pleasewrite} with the alternate boundary conditions \eqref{newBound} to get for $d\neq3$:
\es{stdbbprop3}{
  \Pi_{1,1}^{\zeta}\left(p,z,z'\right)_{i,j}&=\left(zz'\right)^{\frac{d}{2}-1}K_{\frac{d-2}{2}}\left(pz'\right)I_{\frac{2-d}{2}}\left(pz\right)\left(\delta_{i,j}-\frac{p_{i}p_{j}}{p^{2}}\right)\\
  &\qquad +\frac{p_{i}p_{j}}{p^{4}}\left[\left(zz'\right)^{d-1}\partial_{z,z'}\left(\left(zz'\right)^{-d+1}\Pi^{\zeta}_{1,1}(p,z,z')_{z,z'}\right)\right],\\
  \Pi_{1,1}^{\zeta}\left(p,z,z'\right)_{i,z'}&=i\frac{p_{i}}{p^{2}}z^{d-1}\partial_{z}\left(z^{-d+1}\Pi^{\zeta}_{1,1}(p,z,z')_{z,z'}\right),\\
  \Pi_{1,1}^{\zeta}\left(p,z,z'\right)_{z,j}&=-i\frac{p_{j}}{p^{2}}\left(z'\right)^{d-1}\partial_{z'}\left(\left(z'\right)^{-d+1}\Pi^{\zeta}_{1,1}(p,z,z')_{z,z'}\right),\\
  \Pi_{d-1,1}^{\zeta}\left(p,z,z'\right)_{z,z'} &= -\frac{p^{2}}{d-2}\left(\partial_{\Delta}\Pi_{\Delta,0}\left(p,z,z'\right)\right)\mid_{\Delta=d-1} \\
  & \qquad + \frac{4}{(d-2)\Gamma\left(2-\frac{d}{2}\right)\Gamma\left(\frac{d}{2}\right)} \left(\zeta-1\right) \,p^{2}\left(zz'\right)^{\frac{d}{2}}K_{\frac{d-2}{2}}\left(|p|z\right)K_{\frac{d-2}{2}}\left(|p|z'\right)\,,
}
where we fixed the coefficients to match \eqref{newBound},\eqref{newBounds2} and \eqref{Aprop2}.

For $d=3$ and general $k$, we instead get (demanding \eqref{boundF} as boundary conditions)
\es{bb3d}{
 \Pi_{1,1}^{\zeta,k}\left(p,z,z'\right)_{i,j}&=\frac{1}{2p}\Bigg[\frac{e^{-\left|p\right|\left(z'+z\right)}}{\left(\frac{\kappa}{2\pi}\right)^{2}+\left(\frac{1}{16}\right)^{2}}\left(\left(\left(\frac{1}{16}\right)^{2}-\left(\frac{\kappa}{2\pi}\right)^{2}\right)\left(\delta_{i,j}-\frac{p_{i}p_{j}}{p^{2}}\right)-\frac{\kappa}{16\pi}\,\varepsilon_{ijk}\frac{p_{k}}{\left|p\right|}\right)\\&\qquad +e^{-|p|\left|z'-z\right|}\left(\delta_{i,j}-\frac{p_{i}p_{j}}{p^{2}}\right)\Bigg]+\frac{p_{i}p_{j}}{\left|p\right|^{4}}\left[\left(zz'\right)^{2}\partial_{z,z'}\left(\left(zz'\right)^{-2}\Pi_{1,1}^{\zeta,k}\left(p,z,z'\right)_{z,z'}\right)\right]\\\Pi_{1,1}^{\zeta,k}\left(p,z,z'\right)_{i,z'}&=-i\frac{p_{i}}{p^{2}}z^{2}\partial_{z}\left(z^{-2}\Pi_{1,1}^{\zeta,k}\left(p,z,z'\right)_{z,z'}\right)\\\Pi_{1,1}^{\zeta,k}\left(p,z,z'\right)_{z,j}&=i\frac{p_{j}}{p^{2}}\left(z'\right)^{2}\partial_{z'}\left(\left(z'\right)^{-2}\Pi_{1,1}^{\zeta,k}\left(p,z,z'\right)_{z,z'}\right)\\\Pi_{1,1}^{\zeta,k}\left(p,z,z'\right)_{z,z'}&=\frac{\left|p\right|}{2}zz'\Bigg[\left(e^{-\left|p\right|\left(z+z'\right)}\text{Ei}\left(2\left|p\right|z\right)+e^{\left|p\right|\left(z+z'\right)}\text{Ei}\left(-2\left|p\right|z'\right)\right)\\&\qquad\qquad -\left(e^{-\left|p\right|\left(z'-z\right)}\text{Ei}\left(-2\left|p\right|z\right)+e^{-\left|p\right|\left(z-z'\right)}\text{Ei}\left(-2\left|p\right|z'\right)\right)\\&\qquad\qquad +2\frac{1-\zeta}{1+\left(\frac{16\kappa}{2\pi}\right)^{2}}\,e^{-\left|p\right|\left(z+z'\right)}\Bigg]\,.
}

Finally, the difference between $ \Pi_{d-1,1}^{TT}\left(p,z,z'\right)$ and $\Pi_{1,1}^{\zeta,k}\left(p,z,z'\right)$ can be written in terms of the momentum space bulk-to-boundary propagators and the effective photon propagator \eqref{Aprop2Mom} as 
\es{final}{
&{G_{d-1,1}(p,z_1){}_{i|\mu} \langle A_i(p)A_{i'}(-p) \rangle_{\zeta,\infty,k}  
G_{d-1,1}(-p,z_2)}{}_{i'|\mu'}\\
&\qquad=\frac{8\pi^\frac{3d}{2}}{\Gamma\left(\frac{d-2}{2}\right)\Gamma\left(d\right)}
\left(\Pi^{TT}_{d-1,1}(p,z_1,z_2)_{\mu|\mu'}-\Pi^{\zeta,k}_{1,1}(p,z_1,z_2)_{\mu|\mu'}\right)\,,
}
which gives \eqref{identityNew} in position space.

\bibliographystyle{ssg}
\bibliography{ref}
\end{document}